\pdfoutput=1
\documentclass[a4paper,fleqn,usenatbib]{mnras}

\usepackage{graphicx}	% Including figure files
\usepackage{multicol}        % Multi-column entries in tables
\usepackage{url}
\def\aj{AJ}% Astronomical Journal
\def\apj{ApJ}% Astrophysical Journal
\def\apjl{ApJ}% Astrophysical Journal, Letters
\def\apjs{ApJS}% Astrophysical Journal, Supplement
\def\aap{A\&A}% Astronomy and Astrophysics
\def\aaps{A\&AS}% Astronomy and Astrophysics, Supplement
\def\apss{ApSS}% Astrophysics and Space Science
\def\mnras{MNRAS}% Monthly Notices of the RAS
\def\na{NewA}% New Astronomy
\def\pasp{PASP}% 
\def\araa{ARA\&A}%
\usepackage[T1]{fontenc}
\usepackage{ae,aecompl}

\title[Star formation and G/D in ETGs]{The dustier early-type galaxies deviate from late-type galaxies' scaling relations.}
\author[S. Lianou, E. Xilouris, S. C. Madden, and P. Barmby]{S. Lianou$^{1,2}$\thanks{E-mail:
sofia.lianou@cea.fr}, 
E. Xilouris$^{3}$,
S. C. Madden$^{2}$ and
P. Barmby$^{1}$\\
$^{1}$Department of Physics \& Astronomy, University of Western Ontario, London, ON N6A 3K7, Canada\\
$^{2}$Laboratoire AIM, CEA/IRFU/Service d'Astrophysique, Universit\'{e} Paris Diderot, Bat. 709, 91191 Gif-sur-Yvette, France\\
$^{3}$Institute for Astronomy, Astrophysics, Space Applications \& Remote Sensing, National Observatory of Athens, GR-15236 Penteli, Greece}

\date{Accepted XXX. Received YYY; in original form ZZZ}

\pubyear{2016}

\begin{document}
\label{firstpage}
\pagerange{\pageref{firstpage}--\pageref{lastpage}}
\maketitle

\begin{abstract}
Several dedicated surveys focusing on early--type galaxies (ETGs) reveal that significant fractions of them are detectable in all interstellar medium phases studied to date. 
We select ETGs from the \textit{Herschel Reference Survey} that have both far-infrared \textit{Herschel} and either \hbox{H{\sc i}} or \hbox{CO} detection (or both). We derive their star formation rates (SFR), stellar masses and dust masses via modelling their spectral-energy distributions. We combine these with literature information on their atomic and molecular gas properties, in order to relate their star formation, total gas mass and dust mass on global scales. The ETGs deviate from the dust mass--SFR relation and the Schmidt--Kennicutt relation that SDSS star forming galaxies define: compared to SDSS galaxies, ETGs have more dust at the same SFR, or less SFR at the same dust mass. When placing them in the M$_{\star}$-SFR plane, ETGs show a much lower specific SFR as compared to normal star-forming galaxies. ETGs show a large scatter compared to the Schmidt--Kennicutt relation found locally within our Galaxy, extending to lower SFRs and gas mass surface densities. Using an ETG's SFR and the Schmidt--Kennicutt law to predict its gas mass leads to an underestimate. ETGs have similar observed-gas-to-modelled-dust mass ratios to star forming-galaxies of the same stellar mass, as well as they exhibit a similar scatter.
\end{abstract}


%%%%%%%%%%%%%%%%%%%%%%%%%%%%%%%%%%%%%%%%%%%%%%%%%%

%%%%%%%%%%%%%%%%% BODY OF PAPER %%%%%%%%%%%%%%%%%%

\section{Introduction}
\label{sec:introduction}

   Cosmological simulations predict a two-phase galaxy formation scenario: an early epoch where stars formed within the galaxy, and a later epoch where merging and satellite accretion become important in building the mass of the parent galaxy, often forming an early--type galaxy \citep[ETG; e.g.,][]{Oser10,Johansson12,Lackner12}. \citet[]{Thomas10} analyze a large sample of ETGs with SDSS and find the early epoch of star formation to be governed by internal processes, while later phases of star formation activity are influenced by the environment, supporting the two-phase galaxy formation scenario. Additional evidence for a two-phase galaxy formation process comes from observations of globular clusters and metallicity gradients in ETGs \citep[e.g.,][]{Forbes11,Pastorello14}. The occurrence of gravitational interactions and satellite accretion have thus changed the long-standing picture of ETGs being simple stellar systems, passively evolving and devoid of interstellar medium (ISM) with only traces of dust and gas. The importance of minor merging and satellite accretion in driving a significant amount of star formation activity in massive galaxies has been established in \citet{Kaviraj14a,Kaviraj14b}. The star formation and satellite accretion histories in ETGs lead to complex ISM histories. 

   Multiwavelength observational studies have unveiled these complex ISM histories in ETGs, with the detection  of significant amounts of gas and/or dust.
   The detection of dust \citep[e.g.,][]{Xilouris04,Gomez10,Smith12,Rowlands12,Werner14}, and gas \citep[e.g.,][]{Annibali10,Young11,Davis11,Serra12,Kulkarni14}, as well as several stellar tidal features \citep[e.g.,][]{Kim12,Kaviraj12}, lend credence to both an external and an internal origin of the ISM in ETGs. Internal sources are primarily the evolved mass-losing stars that dominate the stellar populations of ETGs polluting the ISM with enriched gas raising their dust abundance \citep{Hirashita15,Bressan06}. External sources are material accreted through gravitational interactions \citep[e.g.,][]{Kaviraj13,Kaneda10}, leading to a contribution of stars, gas and dust to the parent galaxy, probably driving star formation in these systems as studied in \citet{Kaviraj11}. An interplay between an internal origin of the dust found in ETGs from evolved mass-losing stars that formed during an external event, i.e. a past gravitational interaction event, has been suggested for NGC\,4125 \citep{Wilson13}.

   NGC\,5128 is the most prominent nearby example of an ETG with a rich ISM \citep[e.g.,][]{Parkin12}. What is prominent in NGC\,5128 is the gaseous dusty disk, which has been associated with the remnant of an accreted gas--rich galaxy. Moreover, star formation is ongoing in NGC\,5128 in the form of \hbox{H{\sc ii}} regions \citep{Moellenhoff81}. Going beyond NGC\,5128, recent or current star formation activity has been detected directly in nearby ETGs with Hubble Space Telescope and \textit{GALEX} observations \citep[e.g.,][]{Ford13,Yi05}, as well as in the form of \hbox{H{\sc ii}} regions \citep{Bresolin13}, while \textit{Spitzer} IRS \citep{Houck04} spectroscopy has unveiled PAH emission in the dustier ETGs \citep[e.g.,][]{Panuzzo11,Kaneda10,Kaneda07}. The ongoing star formation activity clearly observed in many ETGs together with the richness of the ISM compel us to investigate their relation.
   
   An important physical parameter to constrain chemical evolution models of galaxies is the gas--to--dust (G/D) mass ratio \citep{Galliano08}, which shows the amount of metals locked up in dust. Even though several phases of the ISM of ETGs have been detected, their G/D mass ratio is still an uncertain physical parameter \citep{Wilson13}. Yet, while the bulk of the stars in ETGs were formed in the early Universe, their assembly continues throughout cosmic time \citep{Renzini06}, and understanding how their gas and dust relate provides an important constraint to their evolution. With the availability of \textit{Herschel} observations necessary to constrain the cold dust properties, complemented with surveys focusing on other ISM tracers, this paper aims at exploring the relation among dust, gas and star formation activity in ETGs.
%____________________________________________

\section{Sample \& analysis}
\label{sec:observations}

%
%%%%% TABLE 1 %%%%%%%%%%%%%%%%%%%%%
\begin{table}
     \begin{minipage}[t]{\columnwidth}
       \caption{Global properties for the HRS sub-sample of ETGs studied in this work.}
      \label{sl_table1} 
      \centering
      \renewcommand{\footnoterule}{}
      \begin{tabular}{l c c c c}
\hline\hline
NGC        &HRS          &E(B-V)     &log$_{10}$(M$_{ \hbox{H{\sc i}} }$/M$_{\odot}$)   &log$_{10}$(M$_{H_{2}}$/M$_{\odot}$)\\
(1)        &(2)          &(3)        &(4)                                          &(5)                            \\
\hline
3226       &  3          &0.020      & 8.65                                        &<8.05     \\    
3301       & 14          &0.021      & 7.79                                        &<7.78     \\    
3414       & 22          &0.022      & 8.12                                        &<7.65     \\    
3619       & 45          &0.015      & 8.89                                        &<8.17     \\    
3626       & 46          &0.017      & 9.02                                        &<8.17     \\    
4203       & 93          &0.012      & 9.44                                        & 7.98     \\    
4324       &123          &0.021      & 8.79                                        &<7.78     \\    
4406       &150          &0.026      & 7.95                                        &<7.52     \\    
4429       &161          &0.030      &<7.96                                        & 8.45     \\    
4435       &162          &0.026      &<7.84                                        & 8.34     \\    
4459       &174          &0.040      &<8.39                                        & 8.62     \\    
4469       &176          &0.018      & 7.64                                        & ...      \\    
4526       &200          &0.020      &<7.85                                        & 8.99     \\
4546       &209          &0.029      & 8.25                                        &<7.49     \\
4643       &243          &0.027      & 8.06                                        &<8.16     \\
4684       &253          &0.024      & 8.22                                        &<7.93     \\
4710       &260          &0.026      & 7.76                                        & 8.51     \\
4866       &286          &0.025      & 9.30                                        &<8.48     \\
\hline                                                          
\end{tabular}                                                              
\footnotetext{References for individual galaxies.-- M$_{ \hbox{H{\sc i}}}$ data are adopted from \citet[][their Table 11]{Boselli14a}; M$_{H_{2}}$ data are adopted from \citet[][their Table 12]{Boselli14a}, and the values listed here refer to those that \citet{Boselli14a} computed with a colour--dependent CO-conversion factor.}
\end{minipage}
\end{table}
%%%%%%%%%%%%%%%%%%%%%%%%%%%%%%%%%%%
%
The present galaxy sample is drawn from the \textit{Herschel} Reference Survey \citep[HRS;][]{Boselli10}. This is a volume-limited and K$_{S}$--band flux-limited survey of 323 galaxies in the local universe observed initially with the \textit{Herschel} Space Observatory with all three SPIRE \citep{Griffin10} bands. The SPIRE photometry for the whole HRS galaxy sample is presented by \citet{Ciesla12} and the PACS observations of the HRS were followed up by \citet{Cortese14}. The sample includes galaxies of all morphological types that lie within a distance between 15\,Mpc to 25\,Mpc, and have a 2MASS K$_{S}$--band total magnitude of less than or equal to 12\,mag for the late--type galaxies and 8.7\,mag for the ETGs. The selection criteria have been optimized to study the ISM content of more quiescent galaxies that have a lower limit in the dust mass of $\sim$10$^{4}$\,M$_{\odot}$ \citep[][]{Boselli10}. Many additional observations across the electromagnetic spectrum complement the HRS galaxy sample. For example H$\alpha$ imaging is presented by \citet{Boselli15a}, and SDSS and GALEX imaging by \citet{Cortese12a}. Moreover, HRS galaxies observed in atomic hydrogen and in CO emission have been homogenized and presented by \citet{Boselli14a}.

Thus, the HRS galaxy sample is unique and complete in terms of wavelength coverage, making it well-suited to understand the ISM properties of local galaxies. \citet{Boselli14b,Boselli14c} study the molecular and total gas scaling relations, as well as the effect of the environment on the HRS data set. The present study focuses on the ETG--sample in the HRS with a cold dust component present. There is a total of 62 ETGs in the HRS, first studied by \citet[]{Smith12}. From these, we select those ETGs that have, first, a \textit{Herschel} detection in all three SPIRE bands and, second, either an \hbox{H{\sc i}} detection or a CO detection or both. The SPIRE detection in all three bands is required to examine the cold dust component and derive accurate dust masses. There are 18 ETGs detected in the far-infrared (FIR)/sub-millimeter (submm) with all three SPIRE bands, which also have \hbox{H{\sc i}} or CO: 14 ETGs have \hbox{H{\sc i}} detection, and 6 ETGs are detected in CO, of which 2 have both \hbox{H{\sc i}} and CO detection. These galaxies are listed in Table~\ref{sl_table1}, along with their global properties.

We use all available imaging for these 18 ETGs in order to construct their global multiwavelength spectral energy distributions (SEDs). \citet{Ciesla14} apply the model of \citet{Draine07} to the IR SEDs (from 8$\mu$m to 500$\mu$m) of HRS galaxies, focusing on the sub-sample of gas-rich late-type galaxies. \citet{Cortese12b} also study the dust properties of HRS galaxies and derive dust masses from only three SPIRE bands and employing empirical relations. \citet{Smith12} analyze the dust properties of the 62 ETGs in the HRS using single-temperature modified blackbody fits to their dust SED. \citet{Smith12} use a wavelength coverage ranging from 60$\mu$m to 500$\mu$m, using images for their sample wherever available from IRAS (60 and 100$\mu$m), ISO (60-200$\mu$m; \citealp{Temi04}), \textit{Spitzer} (70-160$\mu$m; \citealp{Bendo12}), PACS (100 and 160$\mu$m; \textit{Herschel} Virgo Cluster Survey \citealp{Davies10}), and SPIRE (250-500$\mu$m; \citealp{Ciesla12}). In our study, we model the full SEDs using two SED models and the same apertures across all wavelengths, in order to derive the star formation and dust properties of the 18 ETGs. Modelling the panchromatic SED using the same extraction aperture throughout the spectrum can provide a consistent constraint to interpret the physical properties of a galaxy \citep{Walcher11}. 

The available data set to construct the SED covers from the ultraviolet (UV) to the FIR/submm regime, and consists of imaging in the GALEX \citep{GilDePaz07}, SDSS \citep{York00}, 2MASS \citep{Skrutskie06}, WISE \citep{Wright10}, \textit{Spitzer} IRAC \citep{Fazio04} and MIPS \citep{Rieke04}, and \textit{Herschel} PACS \citep{Poglitsch10} and SPIRE \citep{Griffin10} bands. The available imaging is drawn from several sources/archives\footnote{http://irsa.ipac.caltech.edu/images.html and http://hedam.lam.fr/HRS/}: from \citet{Cortese12a} for GALEX; the IRAC/WISE/2MASS imaging is accessed through the NASA/ IPAC Infrared Science Archive (IRSA); the MIPS three-band imaging from \citet{Bendo12}; the \textit{Herschel} PACS\,100\,$\mu$m and 160\,$\mu$m photometry are drawn from \citet[][\textit{HIPE} v10.0.0, calibration file v48]{Cortese14}, while the \textit{Herschel} SPIRE images from \citet[][{\sevensize \textit{HIPE}} v8, calibration tree v8.1]{Ciesla12}.

   We process all the available multiwavelength imaging using {\sevensize \textit{IMAGECUBE}} \citep{Lianou13} and performing image registration, convolution to the same point spread function (PSF), and regridding to a common pixel size. In brief, the available imaging is initially converted to a common unit (Jy/pixel) and registered to the same world coordinate system. Then, we background subtract all images and correct the UV--to--MIR imaging for Galactic foreground extinction: 
\begin{equation}
F_{\rm i} = F_{\rm obs} 10^ {0.4 A_{\rm \lambda}},
\end{equation}
where F$_{\rm i}$ is the flux density corrected for foreground extinction, F$_{\rm obs}$ is the observed flux density, and A$_{\rm \lambda}$ is the extinction at the wavelength ${\rm \lambda}$, estimated using the extinction law by \citet{Fitzpatrick99} improved by \citet{Indebetouw05} in the infrared. In the case of the GALEX photometry, we correct for the Galactic foreground extinction using the selective extinction A$_{FUV}/$E(B-V)=8.29 and A$_{NUV}/$E(B-V)=8.18 \citep{Seibert05}. 

%% TABLE 2  %%%%%%%%%%%%%%%%%%%
\begin{table}
      \begin{minipage}[t]{\columnwidth}
        \caption[]{Wavelength coverage and adopted flux calibration uncertainties.
        }
      \label{sl_table2} 
      \renewcommand{\footnoterule}{}
      \begin{tabular}{l c c c}
\hline\hline
Telescope/Filter ($\lambda$;\,$\mu$m)     &FWHM ($\arcsec$) &Uncertainty(\%)  &Ref.\footnote{{\bf References for flux calibration uncertainties.--} (1) \citet{Morrissey07}; (2) \citet{Padmanabhan08}; (3) \citet{Skrutskie06}; (4) \citet{Jarrett11}; (5) \citet{Aniano12}; (6) \citet{Bendo12}; (7) \citet{Remy13}.} \\  
\hline                                  
GALEX/FUV              (0.15)                   &4.2              &5        &1           \\
GALEX/NUV              (0.23)                   &5.3              &3        &1           \\
SDSS/u                 (0.36)                   &1.2              &2        &2           \\
SDSS/g                 (0.46)                   &1.2              &1        &2           \\
SDSS/r                 (0.61)                   &1.2              &1        &2           \\
SDSS/i                 (0.74)                   &1.2              &1        &2           \\
2MASS/J                (1.25)                   &2.5              &3        &3           \\
2MASS/H                (1.65)                   &2.5              &3        &3           \\
2MASS/K$_{S}$           (2.17)                   &2.5              &3        &3           \\
\textit{WISE}/W1          (3.4)                    &8.4              &2.4      &4           \\
\textit{WISE}/W2          (4.6)                    &9.2              &2.8      &4           \\
\textit{WISE}/W3          (12)                     &11.4             &4.5      &4           \\
\textit{WISE}/W4          (22)                     &18.6             &5.7      &4           \\
\textit{Spitzer}/IRAC     (3.6)                    &1.7              &8.3      &5           \\
\textit{Spitzer}/IRAC     (4.5)                    &1.7              &7.1      &5           \\
\textit{Spitzer}/IRAC     (5.8)                    &1.9              &22.1     &5           \\ 
\textit{Spitzer}/IRAC     (8.0)                    &2.0              &16.7     &5           \\ 
\textit{Spitzer}/MIPS     (24)                     &6.0              &4        &6           \\
\textit{Herschel}/PACS    (100)                    &7.1              &5        &7           \\
\textit{Herschel}/PACS    (160)                    &11.2             &5        &7           \\
\textit{Herschel}/SPIRE   (250)                    &18.2             &7        &7           \\
\textit{Herschel}/SPIRE   (350)                    &25.0             &7        &7           \\
\textit{Herschel}/SPIRE   (500)                    &36.4             &7        &7           \\
\hline
\end{tabular} 
\end{minipage}
\end{table}
%%%%%%%%%%%%%%%%%%%%%%%%%%%%%%%%%%%%%%%%%%%%%%%%%%%%%%%%%%%%%%%%%%
%
Subsequently, we convolve all imaging to a common PSF, that of SPIRE\,500$\mu$m, using Gaussian kernels with appropriate full width at half-maximum (FWHM) to match that of SPIRE\,500$\mu$m \citep[43$\arcsec$; Table 6, column 7 in ][]{Aniano11}. Finally, we re--sample the images to the same pixel size of 15$\arcsec$. The uncertainties are estimated from the convolved images, in the same way as described by \citet[see also \citealt{Lianou14} where we recently applied this technique]{Aniano12}. No colour corrections are applied to the \textit{ WISE}, \textit{ Spitzer}, and \textit{ Herschel} images (or any other imaging data set used here), as these are inserted as an additional source of uncertainty. In estimating the photometric uncertainties, we also take into account the calibration uncertainties listed in Table~\ref{sl_table2}. We add all uncertainty components (calibration, background variation, colour correction) in quadrature. The processed imaging is used to construct the global observed SEDs of the sample ETGs, from which the star formation and dust emission properties are derived via SED modelling. We perform photometry in apertures larger than the common PSF used (a FWHM of 43$\arcsec$ for the SPIRE\,500 PSF; \citealt{Aniano11}), in order to derive statistically independent properties. We consistently apply the same aperture to all available imaging, in order to extract the flux densities used to model the ETGs SEDs.
%% TABLE 3  %%%%%%%%%%%%%%%%%%%
\begin{table}
     \begin{minipage}[t]{\columnwidth}
       \caption{Semimajor and semiminor axes, and position angles used to extract photometry in this work, denoted with a, b, and PA, respectively, and in \citet{Ciesla12}, denoted with a$_{C12}$, b$_{C12}$, and PA$_{C12}$, respectively.}
      \label{sl_table3} 
      \centering
      \renewcommand{\footnoterule}{}
      \begin{tabular}{l c c c c c c c}
\hline\hline
%                            & \multicolumn{3}{c}{\sevensize \textit THIS WORK}                            & \multicolumn{3}{c}{\sevensize \textit \citet{Ciesla12}}                 \\
NGC         &HRS         &a        &b        &PA         &a$_{C12}$     &b$_{C12}$         &PA$_{C12}$     \\
(1)         &(2)         &(3)      &(4)      &(5)        &(6)          &(7)              &(8)           \\
\hline                                                           
3226       &  3          &45    &45    &...       &39        &38            &15      \\
3301       & 14          &50    &40    &-45       &42        &42            &55      \\
3414       & 22          &80    &65    &-80       &81        &71            &10      \\
3619       & 45          &80    &75    &-155      &78        &75            &115     \\
3626       & 46          &60    &55    &-150      &58        &54            &160     \\
4203       & 93          &67    &67    &...       &103       &84            &180     \\
4324       &123          &81    &45    &-45       &147       &42            &53      \\
4406       &150          &120   &75    &-40       &129       &113           &130     \\
4429       &161          &80    &65    &-170      &20        &20            &...     \\
4435       &162          &65    &60    &-100      &83        &77            &10      \\
4459       &174          &60    &55    &-10       &20        &20            &...     \\
4469       &176          &90    &65    &-180      &84        &48            &85      \\
4526       &200          &120   &43    &-180      &71        &68            &163      \\
4546       &209          &100   &75    &-180      &104       &75            &80      \\
4643       &243          &115   &110   &-45       &130       &126           &135      \\
4684       &253          &85    &43    &-65       &49        &46            &20      \\
4710       &260          &75    &50    &-50       &103       &60            &27      \\
4866       &286          &113   &60    &-180      &294       &54            &87      \\
\hline                                                                                 
\end{tabular} 
\footnotetext{Note.-- The semimajor and semiminor axes are in units of arcsec, while the position angle in degrees. The minus sign corresponds to position angles measured counter-clockwise from the North.}   
\end{minipage}
\end{table}
%%%%%%%%%%%%%%%%%%%%%%%%%%%%%%%%%%%
%
The apertures used are listed in Table 3.

%%%%% FIGURE 1  %%%%%%%%%% 
 \begin{figure}
  \centering
      \includegraphics[trim=3cm 2.7cm 3cm 3cm,width=8.5cm,clip]{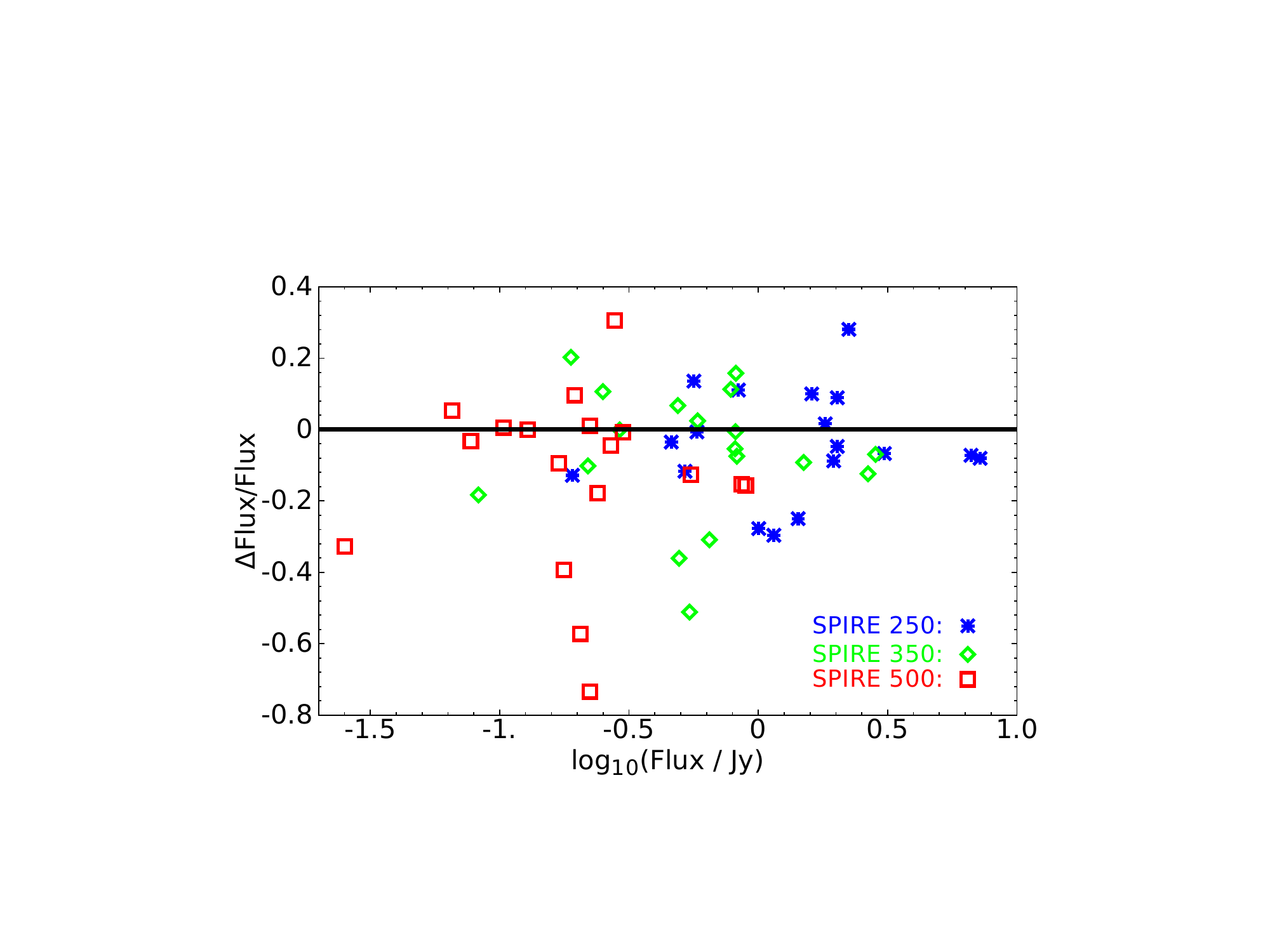}
      \caption{Flux densities from our image processing compared to those of \citet{Ciesla12}. The x--axis shows our flux density, while the y--axis shows the difference between Ciesla's flux density and ours over our flux densities. Blue asterisks correspond to SPIRE\,250 flux densities, green diamonds to SPIRE\,350 flux densities, and red squares to SPIRE\,500 flux densities.
      }
  \label{sl_figure1}
 \end{figure}
%%%%%%%%%%%%%%%%%%%%%%%%%%%%%%%%%%%
%
 Fig.~\ref{sl_figure1} compares the SPIRE flux densities obtained from our image processing with that listed by \citet{Ciesla12}; the latter authors have used the beam areas listed in their Table~1. Hence, for the comparison shown in Fig.~\ref{sl_figure1}, we use the same beam areas as the ones used by \citet{Ciesla12}, while anywhere else in this text we use beam areas 465~arcsec$^{2}$, 823~arcsec$^{2}$, and 1769~arcsec$^{2}$ for SPIRE\,250$\mu$m, 350$\mu$m and 500$\mu$m, respectively\footnote{\url{http://herschel.esac.esa.int/Docs/SPIRE/spire_handbook.pdf}}. For the sake of this comparison, we also use the same photometry apertures as those used by \citet{Ciesla12}, with the exception of HRS\,93, HRS\,123, HRS\,162, HRS\,200, HRS\,209, HRS\,243, HRS\,260, HRS\,286 for which the multiwavelength coverage yields smaller field-of-view, as well as HRS\,161 and HRS\,174 which \citet{Ciesla12} classify as point sources; we have adjusted our apertures to take this into account, and the apertures used are those listed in Table~\ref{sl_table3}. Fig.~\ref{sl_figure1} shows that the fluxes agree within a mean value of 4\%, 7\%, and 13\%, for the SPIRE\,250$\mu$m, SPIRE\,350$\mu$m, and SPIRE\,500$\mu$m, respectively, while the standard deviation is 15\%, 18\%, and 25\%, respectively for each band. The larger standard deviation is contributed by the galaxies which have different apertures than those used by \citet{Ciesla12}. The choice of apertures takes into account the dust emission seen at the SPIRE bands, while in the same time consistently using the same elliptical apertures throughout the whole bands available, which drives our final choice of apertures independently of what is presented by \citet{Ciesla12}.

 The molecular hydrogen masses for the sample are adopted from \citet{Boselli14a} and listed in their Table 1. These authors collect and homogenize for the whole HRS sample all available observations in CO emission, which are adapted either from the literature or from their own observations. To convert the CO emission to molecular hydrogen masses, a conversion factor $X_{\rm CO}$ is applied. \citet{Boselli14a} use both a constant $X_{\rm CO}$ and a luminosity--dependent $X_{\rm CO}$, the latter adapted from \citet{Boselli02}. Here, we choose to show results based on the luminosity--dependent $X_{\rm CO}$; using a constant $X_{\rm CO}$ yields similar trends. A luminosity--dependent $X_{\rm CO}$ is preferred due to the dependence of the $X_{\rm CO}$ on metallicity \citep{Wilson95,Boselli02}, and taking into account the relation between luminosity and metallicity. As most CO surveys target the central part of a galaxy, while CO emission may be found extended beyond this\footnote{NGC\,5128 is the most prominent nearby ETG with molecular gas detected beyond the central parts, in which \citet{Charmandaris00} detect CO at a distance of 15\,kpc from its centre, corresponding to a radius of about 2.7$\times$R$_{eff}$.}, \citet{Boselli14a} have applied a correction for such aperture effects. More specifically, they have assumed a three-dimensional distribution for the CO emission, as has been done in the case of the dust distribution in the z-direction in edge-on galaxies in radiative transfer modelling \citep[e.g.,][]{Xilouris99,deLooze12}. 

 Finally, \citet{Boselli14a} compile from the literature the atomic hydrogen masses for the HRS sample with available measurements, and we use their values here. For those galaxies which are undetected in CO, the M$_{ \hbox{H\,{\sc i}} }$ dominates over their M$_{H_{2}}$, as indicated by the latter's upper limits; hence, when we compute the total gas mass, we include those M$_{H_{2}}$ upper limits. We compute the total gas mass, M$_{gas}$, for each galaxy taking into account the atomic gas mass, M$_{ \hbox{H\,{\sc i}} }$, and the molecular gas mass, M$_{H_{2}}$:
\begin{equation}
M_{\rm gas} = 1.3 (M_{\hbox{H\,{\sc i}}} + M_{H_{2}}).
\end{equation}
The adopted atomic and molecular hydrogen masses are listed in Table~\ref{sl_table1}. To obtain the surface densities of the gas mass, dust mass and star formation rate, we divide these quantities by each galaxy's projected area within the adopted flux extraction aperture. The surface density of a physical property is preferred over its integrated value so as to remove any dependence on galaxy size.
%______________________________________________________________

\section{SED modelling}
\label{sec:modelling}

We derive the star formation and dust emission properties of our ETGs sample via modelling of their SEDs using two models: {\sevensize \textit{PCIGALE}} (v0.9.0)\footnote{\url{http://cigale.lam.fr}} \citep[the python implementation of {\sevensize \textit CIGALE};][]{Roehlly14,Noll09,Burgarella05} and {\sevensize \textit{MAGPHYS}}\footnote{\url{http://www.iap.fr/magphys/magphys/MAGPHYS.html}} \citep{dacunha08}. Both these SED models use the energy balance between the absorption by dust and the re-radiation in the IR to fit the multiwavelength SEDs, from UV to FIR/submm part of the spectrum. The basic components of {\sevensize \textit{MAGPHYS}} are described by \citet{dacunha08}, while {\sevensize \textit{PCIGALE}} is described by \citet[][see \citealt{Noll09} for {\sevensize \textit CIGALE}]{Boquien12}. One important aspect with modelling the SED with {\sevensize \textit{PCIGALE}} is the inclusion of an active galactic nucleus (AGN) component in the SED, while this implementation is currently lacking in {\sevensize \textit{MAGPHYS}}.

{\sevensize \textit{MAGPHYS}} uses the stellar population synthesis of \citet{Bruzual03} to model the star formation histories, assuming both a continuous star formation rate and additional star formation bursts. The attenuation law of \citet{Charlot00} describes the attenuation of the stellar emission by dust in the diffuse ISM and in the stellar birth clouds. No contribution from an AGN to the heating of the ISM is assumed. No detailed modelling of the physical properties of the dust grains is performed in {\sevensize \textit{MAGPHYS}}. While in {\sevensize \textit{MAGPHYS}} pre-computed libraries of SED models already exist, in {\sevensize \textit{PCIGALE}} the user builds the library of the SED models based on input parameters; for a detailed description on input parameters we refer to \citet[][]{Boquien12,Boquien13}.

The assumptions entering the modelling with {\sevensize \textit{PCIGALE}} are as follows. A double exponentially declining star formation history (SFH), with the functional form as in eq.~3 in \citet{Ciesla15}, i.e. $SFR(t) = exp(-t/\tau_{1})$, if $t < t_{1} - t_{2}$, and $SFR(t) = exp(-t/\tau_{1}) + k\ exp(-t/\tau_{2})$, if $t \geq t_{1} - t_{2}$, using the stellar population synthesis models of \citet{Bruzual03}. The optical emission is attenuated in the IR following a modified \citet{Calzetti00} curve, including a UV bump to take into account emission from younger stars as in \citet{Charlot00} \citep{Boquien16}. The dust emission is modelled with the \citet{Draine07} model. An AGN component is assumed for a type--I and a type--II AGN, adopting the model of \citet{Fritz06} which assumes a smooth distribution for the dust around the AGN. Another class of AGN models assumes a clumpy dust distribution, and these two assumptions on the dust distribution can describe several features in the SED of an AGN; a comparison between the two most popular AGN models is given in \citet[][comparing the models of \citealt{Fritz06} and \citealt{Nenkova08}]{Feltre12}.

%
%% TABLE 4  %%%%%%%%%%%%%%%%%%%
\begin{table}
      \begin{minipage}[t]{\columnwidth}
        \caption[]{Input parameter choice for {\sevensize \textit{PCIGALE}}.
        }
      \label{sl_table4} 
      \renewcommand{\footnoterule}{}
      \begin{tabular}{l c}
\hline\hline
                      \multicolumn{2}{c}{Double exponentially declining SFH}      \\
E-folding time of main population (Gyr):  &1.5, 3.0, 4.5, 6.0, 9.0      \\
Age of main stellar population (Gyr):      &13                           \\
E-folding time of late starburst (Myr):   &100, 200, 300, 400, 500      \\
Age of late starburst (Myr):      &100, 200, 300, 500                   \\
Mass fraction of the late burst population: &0, 0.001, 0.01, 0.1        \\
\hline
\multicolumn{2}{c}{\citet{Bruzual03} stellar population synthesis}         \\
Initial mass function:                       &Chabrier                     \\
Metallicity Z:                               &0.02                         \\
\hline
                      \multicolumn{2}{c}{Dust attenuation}                 \\
E(B-V) for the young population:             &0.05, 0.1, 0.2, 0.3, 0.4,    \\
                                             &0.5, 0.75, 1.5, 2.5 \\
Slope $\delta$, \citet{Calzetti00} curve: &-0.5, -0.25, 0 \\
\hline
                      \multicolumn{2}{c}{\citet{Draine07} dust model}      \\
Mass fraction of PAH:                        &0.47, 1.77, 3.90             \\
Minimum radiation field:                     &0.1 -- 25.0                 \\
Maximum radiation field:                     &1000000.0                    \\
Fraction illuminated from Umin to Umax:      &0.0001 -- 0.4                \\
\hline
                      \multicolumn{2}{c}{\citet{Fritz06} AGN model}        \\
Optical depth at 9.7$\mu$m:                  &0.1 \& 10.0                  \\
Full opening angle of the dust torus:        &60                           \\
Angle of equatorial axis and line of sight: &0.001 \& 89.990          \\
AGN fraction:                                &0, 0.05, 0.15        \\
\hline
\end{tabular} 
\end{minipage}
\end{table}
%%%%%%%%%%%%%%%%%%%%%%%%%%%%%%%%%%%%%%%%%%%%%%%%%%%%%%%%%%%%%%%%%%
%
We list in Table~\ref{sl_table4} the non--default input parameters we have adjusted while running {\sevensize \textit{PCIGALE}}. The \citet{Bruzual03} population synthesis model and the \citet{Chabrier03} initial mass function are chosen in {\sevensize \textit{PCIGALE}} so as to be common with {\sevensize \textit{MAGPHYS}}. 

%
%%%%% FIGURE 2  %%%%%%%%%% 
 \begin{figure}
  \centering
      \includegraphics[trim=2cm 6.5cm 2cm 0.5cm,width=8cm,clip]{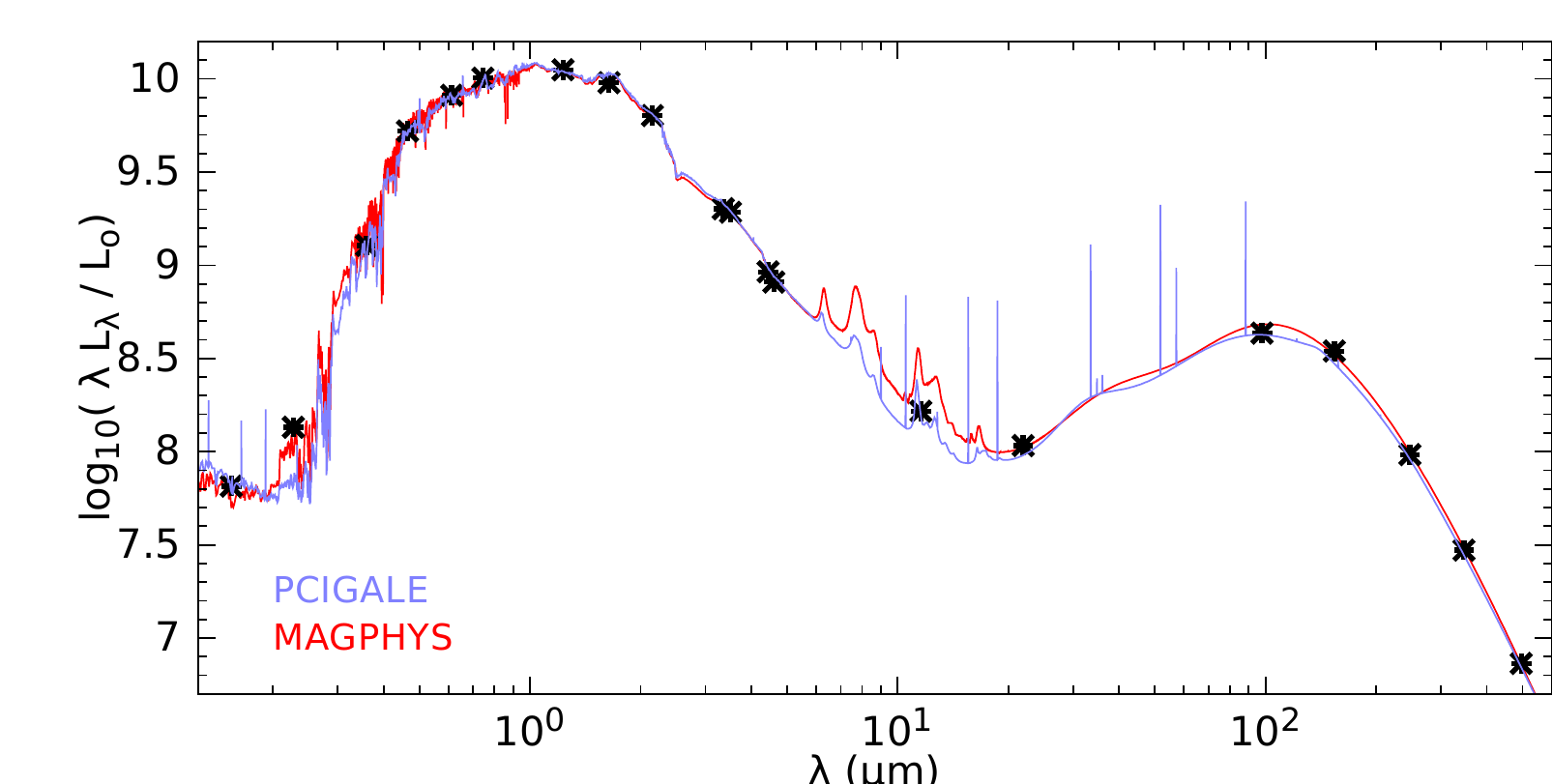}
      \caption{Modelled (best-fitting) and observed SED for NGC\,4203 (HRS\,93) with {\sevensize \textit{PCIGALE}} shown with the blue line, and with {\sevensize \textit{MAGPHYS}} shown with the red line. Black asterisks correspond to observed flux densities, a total of 20 data points used to constrain the SED of NGC\,4203.}
  \label{sl_figure2}
 \end{figure}
%%%%%%%%%%%%%%%%%%%%%%%%%%%%%%%%%%%
%
 Fig.~\ref{sl_figure2} shows an example of a best--fit model SED to an ETG of our sample derived with {\sevensize \textit{PCIGALE}} and {\sevensize \textit{MAGPHYS}}. An important aspect in both SED models is the adoption of a Bayesian approach where probability density functions are constructed for the physical parameters of interest; the median likelihood is adopted in {\sevensize \textit{MAGPHYS}} and the mean value in {\sevensize \textit{PCIGALE}} as the physical parameter value. The 16$^{th}$--84$^{th}$ percentile range is then used to reflect the associated confidence interval in {\sevensize \textit{MAGPHYS}} \citep{dacunha08}, and the standard deviation in {\sevensize \textit{PCIGALE}} \citep{Noll09}. We adopt in this study the Bayesian approach built in the SED models, rather than the physical parameters inherent to the best--fit SED model. We consider the parameters of the mass of the dust (M$_{D}$), the star formation rate (SFR) averaged over the last 10$^{8}$yr, and the stellar mass (M$_{\star}$). We adopt the following naming convention throughout the remaining text: we use a subscript ``P'' to denote results derived with {\sevensize \textit{PCIGALE}}, while a subscript ``M'' is used to denote results with {\sevensize \textit{MAGPHYS}}.

 As the SED model libraries in either {\sevensize \textit{MAGPHYS}} or {\sevensize \textit{PCIGALE}} are built based on different assumptions for the input parameters, e.g. SFH or dust model assumption or attenuation law, the derived properties with the two modelling techniques are not directly comparable. Each of the two SED models defines a unique ''scale'' for the derived properties. The use of ``property scale'' is the same in concept as that of metallicity scale: the value of the metallicity (or property) is tied to the method used to derive it; for example, the \citet{Zinn84} or the \citet{Carretta97} metallicity scale for stellar metallicities derived using the Ca\,{\sc ii} triplet, as well as the \citet{Pilyugin05} metallicity scale for gas--phase metallicities.

 Hence, any comparison of the derived properties between the two SED models performed in this study is only illustrative and has the purpose of understanding the intrinsic difference between the two scales, rather than to derive a calibration between them. The reason for this is that a calibration between the two SED model scales is not unique, as it depends on the input parameters chosen to build each SED library used in the fitting process. Hence, deriving a calibration requires a complete search of the available parameter space of the input parameters in the SED models; this goes beyond the scope of this paper. The same reasoning holds for the comparison of the derived properties between each of the two SED models and those assuming a modified blackbody.
 
%%%%% FIGURE 3  %%%%%%%%%% 
 \begin{figure}
  \centering
      \includegraphics[trim=2.5cm 2cm 4cm 0.5cm,width=8.5cm,clip]{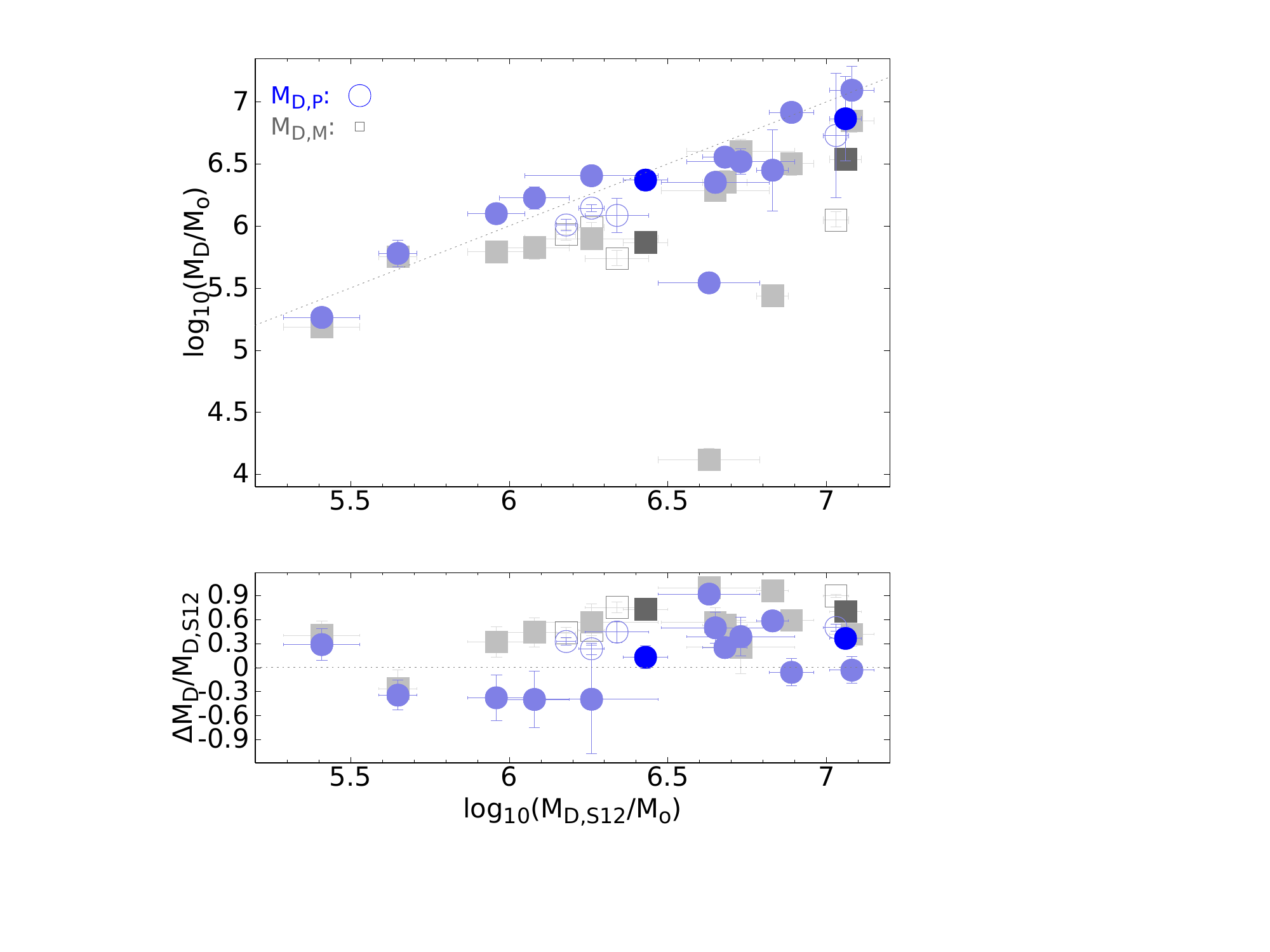}
      \caption{Comparison between M$_{D}$ derived in this work to that derived by \citet{Smith12} (upper panel), along with their relative residual, i.e. M$_{D,S12}$ minus M$_{D}$ over M$_{D,S12}$ (lower panel).
 {\sevensize \textit{MAGPHYS}} results are shown in grey squares, while {\sevensize \textit{PCIGALE}} results in blue circles. Dark-coloured filled symbols indicate ETGs with both \hbox{H\,{\sc i}} and CO; light-coloured filled symbols are ETGs with only \hbox{H\,{\sc i}}; open symbols are ETGs with only CO. The solid line in the upper panel is the one-to-one relation.}
 \label{sl_figure3}
 \end{figure}
%%%%%%%%%%%%%%%%%%%%%%%%%%%%%%%%%%%
%
 Fig.~\ref{sl_figure3} illustrates the difference in M$_{D}$ derived here, grey squares for {\sevensize \textit{MAGPHYS}} and blue circles for {\sevensize \textit{PCIGALE}}, with that derived by \citet{Smith12} using modified blackbody fits to the IR/submm SED (60--to--500\,$\mu$m), denoted as M$_{D,S12}$ and shown in the upper panel. The lower panel shows the relative residuals of the comparison. \citet{Smith12} use the SPIRE photometry from \citet{Ciesla12} with beam areas as listed in Table~1 of the latter. Hence, for the comparison between Smith's and our dust masses (both with {\sevensize \textit{MAGPHYS}} and {\sevensize \textit{PCIGALE}}) we have used the same apertures to extract fluxes, as well as the same beam areas as by \citet{Ciesla12} (see also explanation for Fig.~\ref{sl_figure1} in previous section). In this comparison shown in  Fig.~\ref{sl_figure3}, the mean dust mass from \citet{Smith12} is log$_{10}$(M$_{D,S12})$=6.5$\pm$0.5, while the mean dust mass with {\sevensize \textit{MAGPHYS}} is log$_{10}$(M$_{D,M})$=6.0$\pm$0.6 and with {\sevensize \textit{PCIGALE}} is log$_{10}$(M$_{D,P})$=6.3$\pm$0.5. The mean relative residual of the comparison shown in the lower panel of Fig.~\ref{sl_figure3} is 51\%, with a standard deviation of 28\% for {\sevensize \textit{MAGPHYS}}, while in the case of {\sevensize \textit{PCIGALE}} the mean relative residual is 18\%  with a standard deviation of 38\%.

 Fig.~\ref{sl_figure3} shows a systematic tendency of dust masses derived with {\sevensize \textit{MAGPHYS}} to be underestimated as compared to dust masses derived with modified blackbody fits; this seems to be random in the case of {\sevensize \textit{PCIGALE}}. \citet{Appleton14} study the star formation and dust properties in NGC\,3226; they find that dust masses derived with {\sevensize \textit{MAGPHYS}} are underestimated by 30\% as compared to dust masses derived with modified blackbody fits. Their dust mass estimate with {\sevensize \textit{MAGPHYS}} on NGC\,3226 (HRS\,3), equal to 6.09 in logarithmic space, is consistent with our finding of 5.8$\pm$0.1 in logarithmic space, considering that we probe a slightly different area. 

%%%%% FIGURE 4  %%%%%%%%%% 
 \begin{figure}
  \centering
      \includegraphics[trim=2cm 6.5cm 4cm 0.5cm,width=8.5cm,clip]{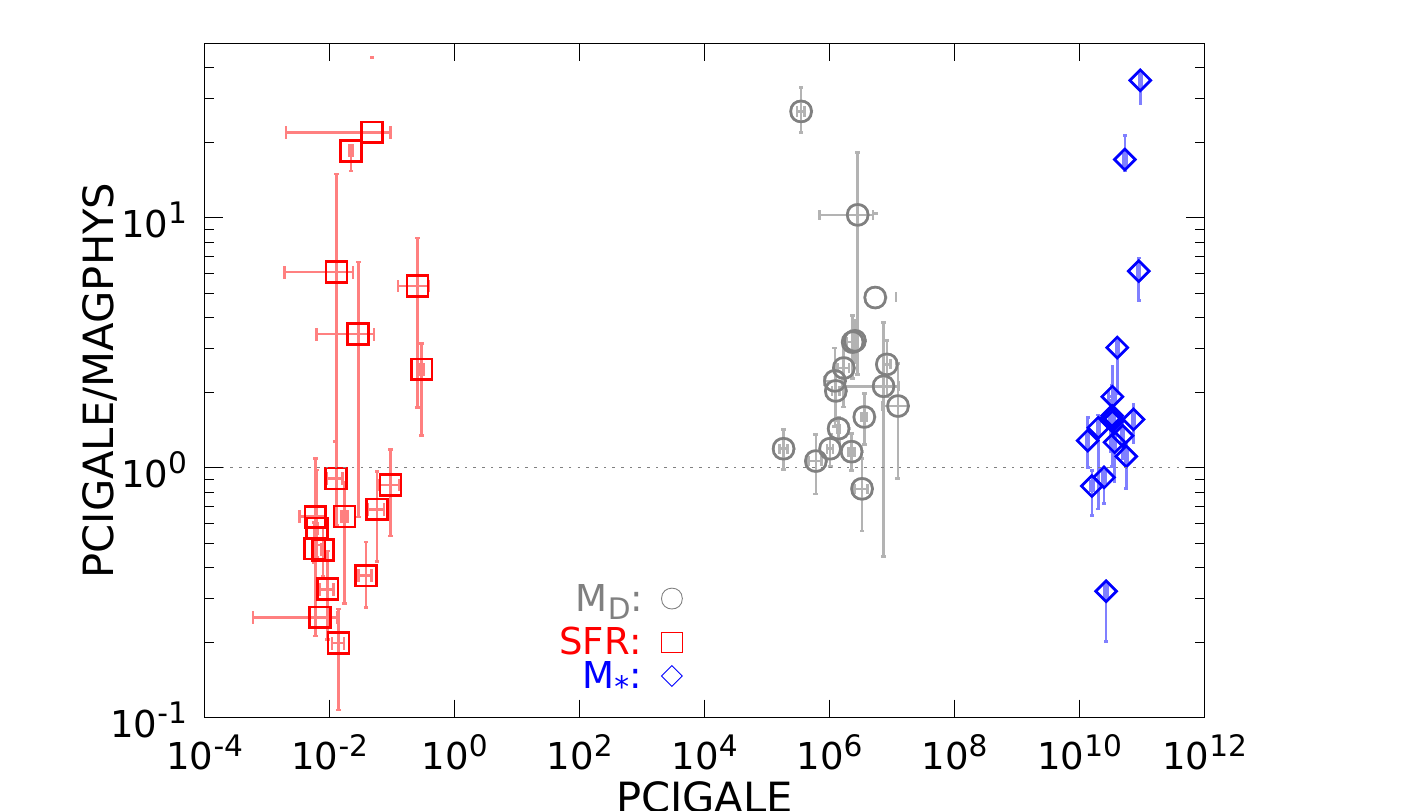}
      \caption{Comparison of the physical property derived with the two SED models: ratio of {\sevensize \textit{MAGPHYS}} property over {\sevensize \textit{PCIGALE}} property as a function of the {\sevensize \textit{PCIGALE}} property. The physical properties compared are: M$_{D}$ (in M$_{\odot}$) shown with grey circles, SFR (in M$_{\odot}$yr$^{-1}$) shown with red squares, M$_{\star}$ (in M$_{\odot}$) shown with blue diamonds.}
 \label{sl_figure4}
 \end{figure}
%%%%%%%%%%%%%%%%%%%%%%%%%%%%%%%%%%%
%   
 Fig.~\ref{sl_figure4} illustrates the difference in the derived properties between {\sevensize \textit{PCIGALE}} and {\sevensize \textit{MAGPHYS}}: M$_{D}$ (in M$_{\odot}$) shown with grey circles, SFR (in M$_{\odot}$yr$^{-1}$) shown with red squares, M$_{\star}$ (in M$_{\odot}$) shown with blue diamonds. Note that this comparison is performed on beam areas and apertures used in this paper (see Table~\ref{sl_table3}). The dust masses derived with {\sevensize \textit{MAGPHYS}} are underestimated as compared to those derived with {\sevensize \textit{PCIGALE}}, by a factor of 4$\pm$6 on average. The comparison of the SFRs derived with the two models shows a large scatter with a variation of the  {\sevensize \textit{MAGPHYS}} SFR between 0.02 to 120 times that of {\sevensize \textit{PCIGALE}}. The stellar masses derived with {\sevensize \textit{MAGPHYS}} are underestimated as compared to those derived with {\sevensize \textit{PCIGALE}} by a factor of 4.4$\pm$8.6 on average. Overall, as stated earlier, the two SED models can be regarded each as defining a separate scale for each physical parameter, with model SED libraries built based on different assumptions, hence they are expected to have differences in the derived parameters.

 \citet{Ciesla15} use {\sevensize \textit CIGALE} on mock galaxies, in order to investigate the influence of an AGN component in modelling their SEDs; they find that including an AGN component in the SED modelling is important to constrain the derived physical properties.  While {\sevensize \textit{PCIGALE}} includes an AGN component in the SED, this is not currently the case in {\sevensize \textit{MAGPHYS}}.  There are two ETGs in our sample hosting weak AGN activity \citep[HRS\,3 and HRS\,93;][]{Appleton14,Smith12}. \citet{Appleton14} suggest the likely presence of a low-luminosity AGN contribution to the IR for NGC\,3226, after modelling its SED using two different methods each including an AGN component; they constrain the AGN contribution to the IR luminosity to be between 20$\pm$5\% to 18$\pm$4\%. Table~5 by \citet{Appleton14} suggests that the dust mass of NGC\,3226 derived when using {\sevensize \textit{MAGPHYS}} can be underestimated by a factor of 3 as compared to the dust mass derived using one of the SED modelling method that includes an AGN component. 

 As stated earlier, in addition to the inclusion or not of an AGN component, different modelling techniques can lead to different dust mass estimates. This is the conclusion reached in studies where detailed dust grain modelling versus modified blackbody fits are compared \citep{Dale12,Magdis12}. To this end, \citet{Berta16} discuss, as \citet{Bianchi13} also points out, that adopting the same mass absorption coefficient of dust ($\kappa_{\nu}$) and the same dust emissivity index ($\beta$) is important in deriving consistent dust masses when comparing different dust modelling techniques. In this aspect, absolute dust masses, and direct comparison between dust mass estimates derived from different modelling techniques, are not as meaningful as is the examination of trends in scaling relations within each of the SED scales, as these are defined earlier in this section.

%
%%%%% TABLE 5 %%%%%%%%%%%%%%%%%%%%%
\begin{table*}
     \begin{minipage}[t]{\textwidth}
      \caption{Derived properties obtained with {\sevensize \textit{MAGPHYS}, indicated with an ''M'' subscript, and with {\sevensize \textit{PCIGALE}}, indicated with a ''P'' subscript, for the ETG sample.}}
      \label{sl_table5} 
      \centering
      \renewcommand{\footnoterule}{}
      \begin{tabular}{l c c c c c c c}
\hline\hline
%                            & \multicolumn{3}{c}{\sevensize \textit MAGPHYS OUTPUT}                            &\multicolumn{3}{c}{\sevensize \textit PCIGALE OUTPUT}                 \\
NGC         &HRS         &log$_{10}$(SFR$_M$)      &log$_{10}$(M$_{\star,M}$)  &$log_{10}$(M$_{D,M}$)        &log$_{10}$(SFR$_{P}$)    &log$_{10}$(M$_{\star,P}$)   &log$_{10}$(M$_{D,P}$)  \\
(1)         &(2)         &(3)                     &(4)                     &(5)                        &(6)                     &(7)                     &(8)                  \\
\hline                                                                                                                                                             
3226       &  3          &-1.56$^{+0.10}_{0.23}$     &10.02$^{+0.10}_{-0.09}$    &5.79$^{+0.08}_{-0.09}$        &-2.16$\pm$0.40     	&10.13$\pm$0.02	     	 &6.10$\pm$0.06      \\
3301       & 14          &-2.03$^{+0.24}_{0.22}$     &10.14$^{+0.05}_{-0.23}$    &5.19$^{+0.06}_{-0.04}$        &-2.23$\pm$0.19     	&10.30$\pm$0.02	     	 &5.26$\pm$0.06      \\
3414       & 22          &-1.55$^{+0.15}_{0.12}$     &10.46$^{+0.11}_{-0.13}$    &5.90$^{+0.09}_{-0.08}$        &-2.03$\pm$0.11     	&10.56$\pm$0.02	     	 &6.41$\pm$0.02      \\
3619       & 45          &-1.08$^{+0.13}_{0.11}$     &10.30$^{+0.08}_{-0.11}$    &6.50$^{+0.09}_{-0.10}$        &-1.24$\pm$0.13     	&10.50$\pm$0.02	       	 &6.92$\pm$0.06      \\
3626       & 46          &-0.92$^{+0.11}_{0.20}$     &10.31$^{+0.14}_{-0.16}$    &6.29$^{+0.07}_{-0.05}$        &-0.53$\pm$0.04     	&10.52$\pm$0.02	     	 &6.35$\pm$0.05      \\
4203       & 93          &-1.78$^{+0.10}_{0.10}$     &10.34$^{+0.12}_{-0.10}$    &5.87$^{+0.09}_{-0.09}$        &-2.11$\pm$0.02     	&10.53$\pm$0.02	     	 &6.37$\pm$0.08      \\
4324       &123          &-0.99$^{+0.12}_{0.04}$     &10.27$^{+0.06}_{-0.10}$    &6.35$^{+0.10}_{-0.09}$        &-1.42$\pm$0.10     	&10.20$\pm$0.03	     	 &6.56$\pm$0.05      \\
4406       &150          &-2.93$^{+0.01}_{0.07}$     & 9.42$^{+0.02}_{-0.08}$    &4.12$^{+0.09}_{-0.05}$        &-1.66$\pm$0.02     	&10.97$\pm$0.02	     	 &5.54$\pm$0.06      \\
4429       &161          &-1.57$^{+0.16}_{0.24}$     &10.67$^{+0.06}_{-0.08}$    &5.99$^{+0.04}_{-0.04}$        &-1.76$\pm$0.05     	&10.86$\pm$0.02	     	 &6.14$\pm$0.03      \\
4435       &162          &-2.67$^{+0.52}_{0.13}$     &10.12$^{+0.03}_{-0.15}$    &5.74$^{+0.07}_{-0.06}$        &-1.89$\pm$0.37     	&10.60$\pm$0.02	     	 &6.09$\pm$0.14      \\
4459       &174          &-1.86$^{+0.13}_{0.44}$     &10.56$^{+0.05}_{-0.08}$    &5.93$^{+0.04}_{-0.05}$        &-1.90$\pm$0.12     	&10.69$\pm$0.02	     	 &6.01$\pm$0.05      \\
4469       &176          &-2.66$^{+0.11}_{0.23}$     & 9.49$^{+0.10}_{-0.04}$    &5.44$^{+0.08}_{-0.07}$        &-1.32$\pm$0.42     	&10.72$\pm$0.03	     	 &6.45$\pm$0.33      \\
4526       &200          &-2.08$^{+0.23}_{0.09}$     &10.16$^{+0.05}_{-0.10}$    &6.05$^{+0.07}_{-0.06}$        &-1.54$\pm$0.34     	&10.95$\pm$0.02	     	 &6.73$\pm$0.50      \\
4546       &209          &-1.92$^{+0.12}_{0.05}$     &10.43$^{+0.01}_{-0.09}$    &5.83$^{+0.09}_{-0.10}$        &-2.24$\pm$0.02     	&10.39$\pm$0.02	     	 &6.23$\pm$0.09      \\
4643       &243          &-1.16$^{+0.13}_{0.18}$     &10.70$^{+0.05}_{-0.11}$    &6.85$^{+0.08}_{-0.09}$        &-1.86$\pm$0.10     	&10.75$\pm$0.02	     	 &7.09$\pm$0.19      \\
4684       &253          &-0.96$^{+0.07}_{0.06}$     &10.32$^{+0.02}_{-0.11}$    &5.75$^{+0.06}_{-0.04}$        &-1.03$\pm$0.15     	&10.52$\pm$0.02	     	 &5.78$\pm$0.11      \\
4710       &260          &-1.32$^{+0.09}_{0.19}$     &10.24$^{+0.13}_{-0.18}$    &6.54$^{+0.07}_{-0.06}$        &-0.60$\pm$0.22     	&10.52$\pm$0.06	     	 &6.86$\pm$0.34      \\
4866       &286          &-1.96$^{+0.31}_{0.10}$     &10.92$^{+0.02}_{-0.16}$    &6.60$^{+0.10}_{-0.10}$        &-2.21$\pm$0.02     	&10.43$\pm$0.02	     	 &6.52$\pm$0.10      \\
\hline                                                                                 
\end{tabular} 
\footnotetext{Note.-- The unit for SFR is M$_{\odot}$yr$^{-1}$, while for M$_{\star}$ and M$_{D}$ is M$_{\odot}$.}   
\end{minipage}
\end{table*}
%%%%%%%%%%%%%%%%%%%%%%%%%%%%%%%%%%%
%
Table~\ref{sl_table5} lists the results obtained with {\sevensize \textit{PCIGALE}} and {\sevensize \textit{MAGPHYS}}, along with their associated uncertainties, following the Bayesian approach. For these 18 ETGs, the mean and standard deviation of the derived properties are:
log$_{10}$(M$_{D,M})$\,=\,5.9$\pm$0.6 and log$_{10}$(M$_{D,P})$\,=\,6.3$\pm$0.5, log$_{10}$(M$_{\star,M}$)\,=\,10.3$\pm$0.4 and log$_{10}$(M$_{\star,P}$)\,=\,10.6$\pm$0.2 (both M$_{D}$ and M$_{\star}$ in M$_{\odot}$), log$_{10}$(SFR$_{M}$)\,=\,$-$1.7$\pm$0.6 and log$_{10}$(SFR$_{P}$)\,=\,$-$1.7$\pm$0.6 (M$_{\odot}$yr$^{-1}$). Overall, while taking the mean value and standard deviation of each property indicates consistency between values derived based on each of the two SED models, nevertheless, the comparison shown in Fig.~\ref{sl_figure4} shows large one-to-one differences among the properties, as discussed earlier.
%__________________________________________________________________

\section{Dust, gas, and star formation}

%
%%%%% FIGURE 5  %%%%%%%%%% 
 \begin{figure}
  \centering
      \includegraphics[trim=3cm 2.5cm 3cm 3cm,width=8.5cm,clip]{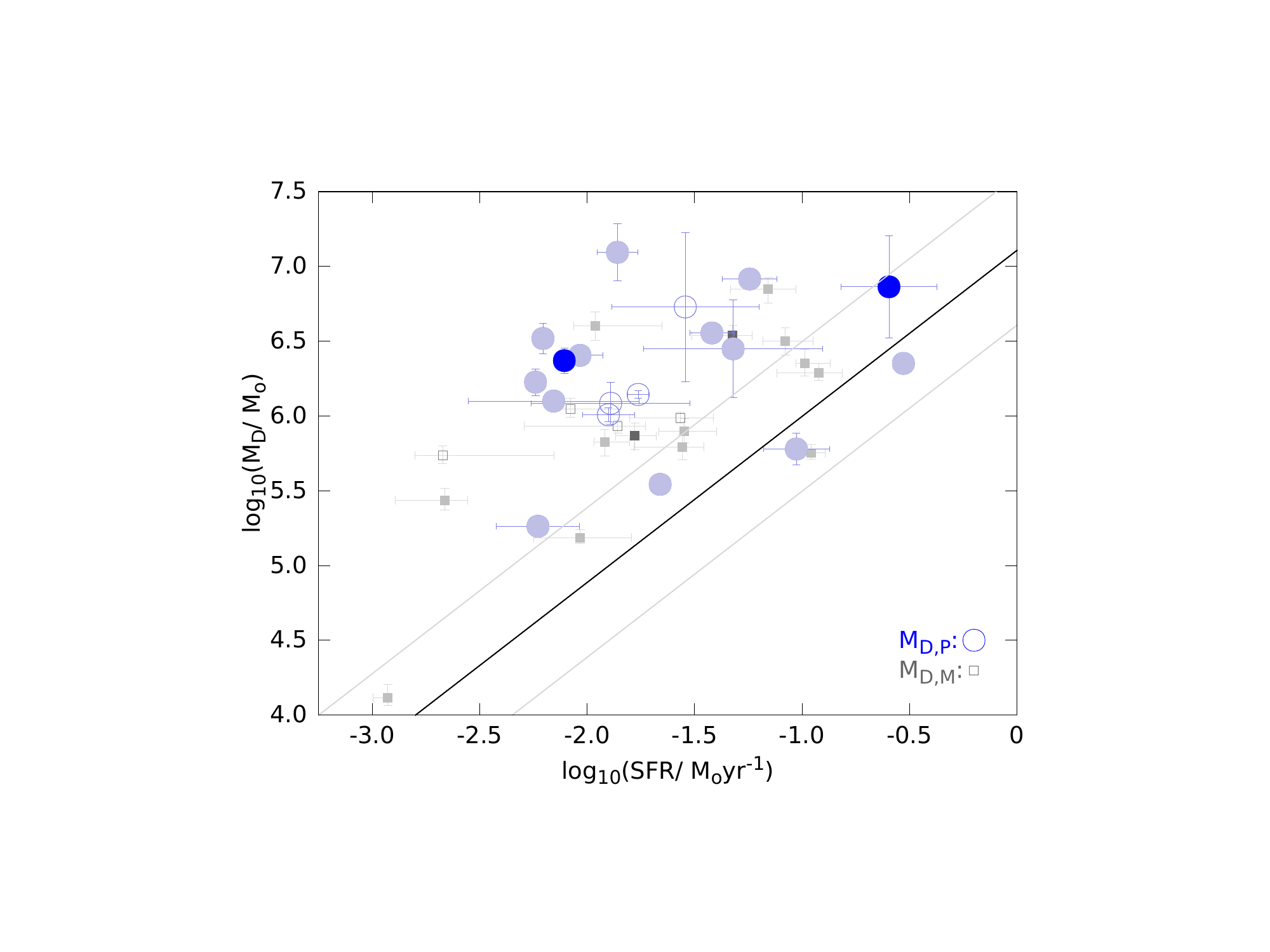}
      \caption{Dust mass, M$_{D}$ as a function of the star formation rate, SFR.
     {\sevensize \textit{MAGPHYS}} results are shown in grey squares, while {\sevensize \textit{PCIGALE}} results in blue circles. Dark-coloured filled symbols indicate ETGs with both \hbox{H\,{\sc i}} and CO; light-coloured filled symbols are ETGs with only \hbox{H\,{\sc i}}; open symbols are ETGs with only CO.
        The black solid line is the relation found by \citet{dacunha10b} for a large number of star-forming galaxies, while the two thin grey-lines represent the intrinsic scatter of this relation. The error bars represent the 16$^{th}$--84$^{th}$ percentile range for {\sevensize \textit{MAGPHYS}} and the standard deviation of the probability distribution function for {\sevensize \textit{PCIGALE}}.
  }
  \label{sl_figure5}
 \end{figure}
%%%%%%%%%%%%%%%%%%%%%%%%%%%%%%%%%%%
%
 If we attribute the cold dust emission of ETGs as due to recent star formation activity, then we expect the dust mass and SFR to be tracing each other. Fig.~\ref{sl_figure5} shows the dust mass as a function of the SFR, with dust masses derived with {\sevensize \textit{MAGPHYS}} shown in grey squares and with {\sevensize \textit{PCIGALE}} in blue circles. \citet{dacunha10b} study a sample of low--redshift galaxies drawn from the SDSS DR6 main spectroscopic sample with additional GALEX, 2MASS and IRAS wavelength coverage. These authors find that the galaxies in their sample follow a tight correlation between dust mass and SFR averaged over the last 10$^{8}$\,yr; they suggest this relation as a calibration to derive dust masses of galaxies given their SFRs. The relation of \citet[][their eq.~9.]{dacunha10b} is shown in Fig.~\ref{sl_figure5} with the black solid line, while the intrinsic scatter is shown with the grey solid lines. In the case of the ETGs, Fig.~\ref{sl_figure5} shows that these deviate from this tight relation: the majority of the ETGs are scattered beyond the intrinsic scatter of the relation introduced by \citet{dacunha10b}. In addition, ETGs fall into the lower left regime of Fig.~5 in \citet{dacunha10b}, i.e. towards lower dust masses and lower SFRs.

 There may be several factors contributing to this scatter seen for the ETGs in Fig.~\ref{sl_figure5}. First, the dust emission is not solely due to recent star formation, but can also be attributed to more evolved stars. Such dust emission, though, is not adequate to justify the amount of dust mass detected in these ETGs \citep[][and references therein]{Kulkarni14}. Then, the interstellar medium in these galaxies may have been externally acquired, via mergers and satellite accretion. This latter suggestion is in line with the study of \citet{Clemens10}, who find short grain lifetimes in passive elliptical galaxies, assuming their dust mass is solely due to evolved stars. \citet{Clemens10} suggest that in the case of FIR/submm-detected ETGs the dust mass should be externally acquired. By selection, our ETGs fall into this category.

%
%%%%% FIGURE 5  %%%%%%%%%% 
 \begin{figure}
  \centering
      \includegraphics[trim=3cm 2.7cm 3cm 3cm,width=8.5cm,clip]{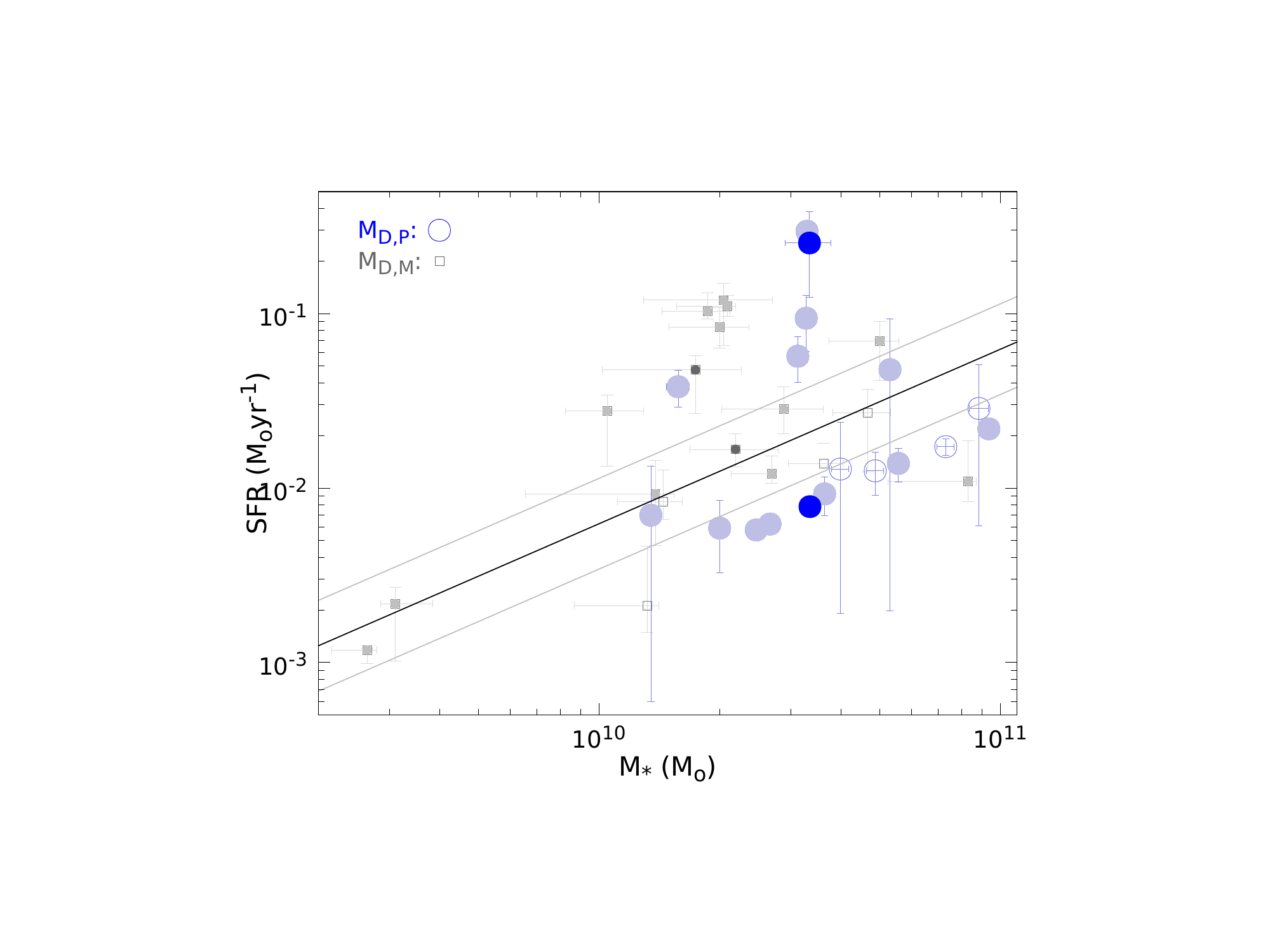}
      \caption{Star formation rate, SFR, as a function of stellar mass, M$_{\star}$. The solid black line indicates the relation of local galaxies studied in \citet[][their figure 16]{Elbaz11} scaled down by a factor of 400, while the grey lines indicate the 0.26\,dex rms of the relation. 
        {\sevensize \textit{MAGPHYS}} results are shown in grey squares, while {\sevensize \textit{PCIGALE}} results in blue circles. Dark-coloured filled symbols indicate ETGs with both \hbox{H\,{\sc i}} and CO; light-coloured filled symbols are ETGs with only \hbox{H\,{\sc i}}; open symbols are ETGs with only CO.
        The error bars represent the 16$^{th}$--84$^{th}$ percentile range for {\sevensize \textit{MAGPHYS}} and the standard deviation of the probability distribution function for {\sevensize \textit{PCIGALE}}.
  }
  \label{sl_figure5b}
 \end{figure}
%%%%%%%%%%%%%%%%%%%%%%%%%%%%%%%%%%%
% 
     Fig.~\ref{sl_figure5b} places the ETGs in the SFR versus M$_{\star}$ plane. Normal star-forming galaxies are found to follow a tight correlation in this plane \citep[][and references therein]{Brinchmann04,Elbaz07,Noeske07}. The slope of this relation is consistent with $\sim$one independent of redshift, but the normalization changes such that at higher redshift the star formation is higher for a given mass \citep{Elbaz11}. The solid line in Fig.~\ref{sl_figure5b} represents a scaled-down (by 400) version of the relation defined for local star forming galaxies studied by \citet[][their figure 16, left panel]{Elbaz11}, i.e.~for the scaled down relation SFR\,$\propto$\,M$_{\star}$/[400$\times$4$\times$10$^{9}$M$_{\odot}$]; the scaling-down has been performed to facilitate the presentation of the figure. The grey lines in Fig.~\ref{sl_figure5b} show the 0.26\,dex rms scatter of the original relation presented in \citep{Elbaz11}. It is remarkable that this scatter in SFR remains the same across all redshift, suggesting that stellar feedback processes likely play a role to its constancy \citep{Schreiber15}. Fig.~\ref{sl_figure5b} shows that ETGs scatter around the scaled-down version of the relation of \citet{Elbaz11}, i.e. ETGs fall systematically below the original relation found for local star-forming galaxies by \citet{Elbaz11}, having lower SFRs for a given mass as compared to normal star-forming galaxies. In other words, while normal star-forming galaxies in the SFR-M$_{\star}$ plane have a constant specific SFR, defined as the ratio of SFR with stellar mass, of 0.25Gyr$^{-1}$ \citep{Elbaz11}, the ETGs in this study have specific SFRs lower by a factor of 30 to 4000 than that of normal galaxies.

%%%%% FIGURE 6  %%%%%%%%%% 
 \begin{figure}
  \centering
      \includegraphics[trim=3cm 2.7cm 3cm 3cm,width=8.5cm,clip]{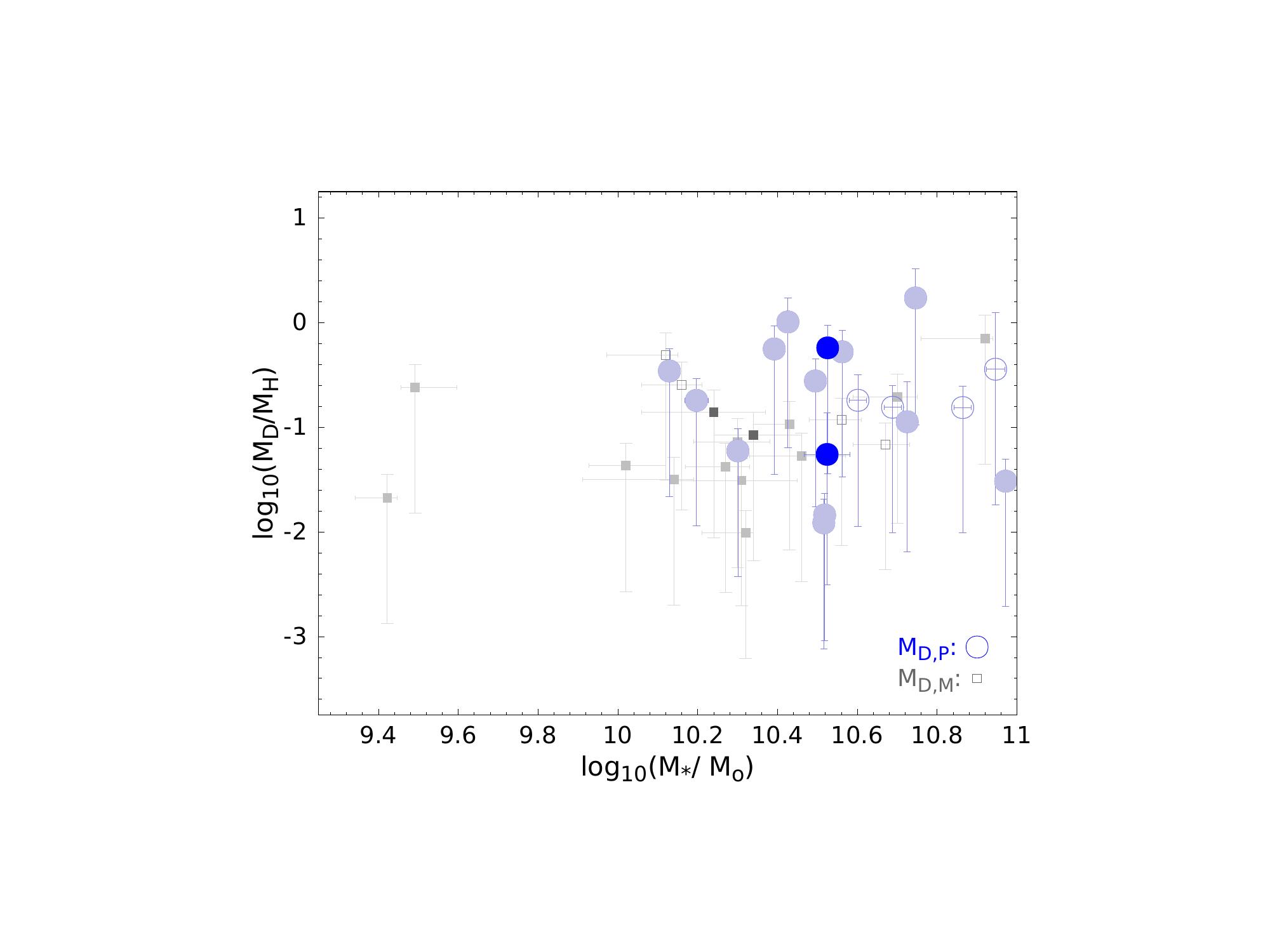}
      \includegraphics[trim=3cm 2.7cm 3cm 3cm,width=8.5cm,clip]{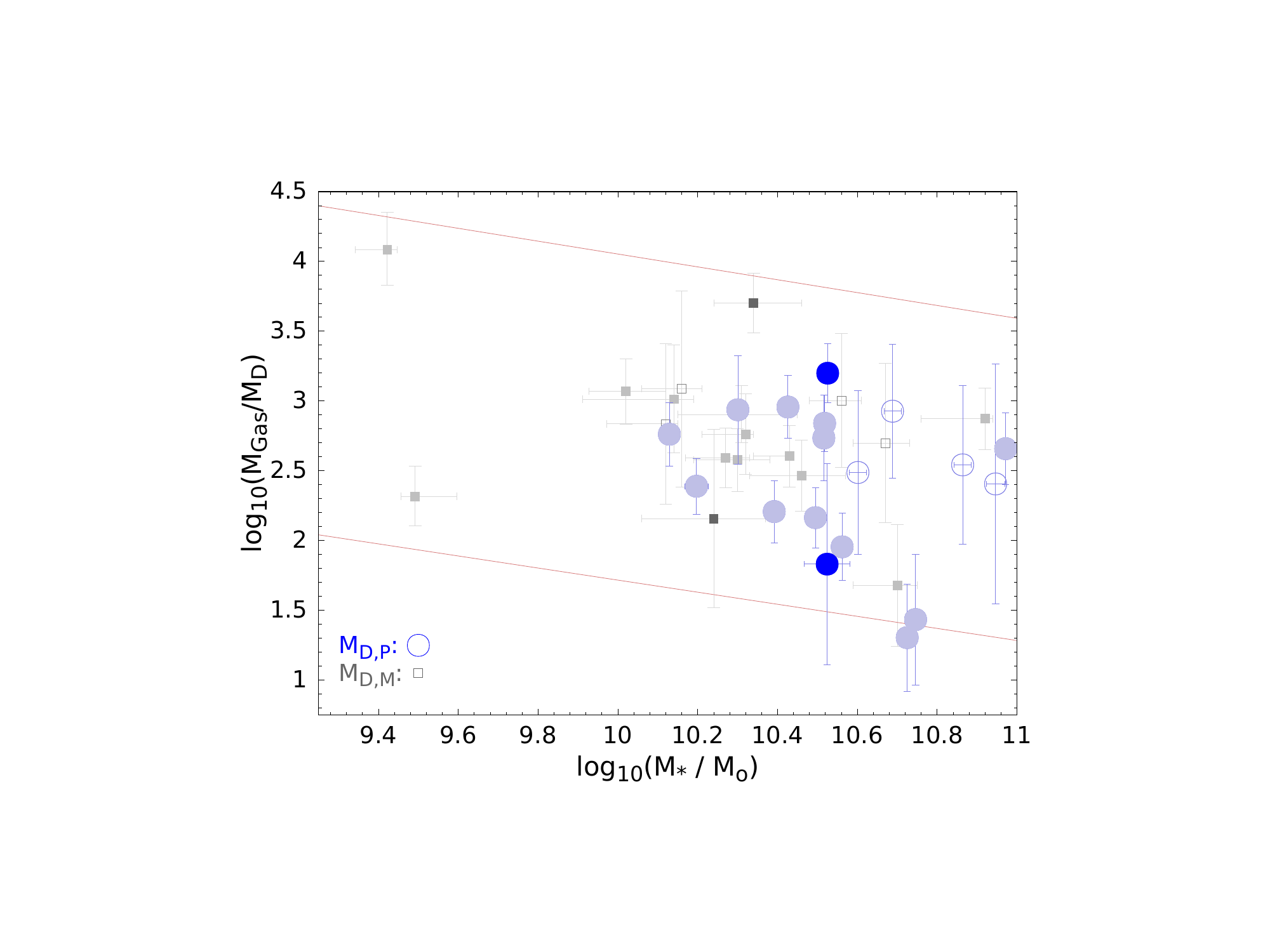}
  \caption{\textit{ Upper panel}: D/G mass ratio, M$_{D}$/ M$_{H}$ as a function of the stellar mass, M$_{\star}$, assuming a Schmidt-Kennicutt law to derive total gas masses from the SFR. 
    \textit{ Lower panel}: G/D mass ratio, M$_{Gas}$/M$_{D}$, using observed total gas masses. The solid red lines indicate the range of values found for the G/D mass ratio as a function of M$_{\star}$ by \citet{Remy14}.
    In both panels, {\sevensize \textit{MAGPHYS}} results are shown in grey squares, while {\sevensize \textit{PCIGALE}} results in blue circles. Dark-coloured filled symbols indicate ETGs with both \hbox{H\,{\sc i}} and CO; light-coloured filled symbols are ETGs with only \hbox{H\,{\sc i}}; open symbols are ETGs with only CO. 
    The error bars represent the 16$^{th}$--84$^{th}$ percentile range for {\sevensize \textit{MAGPHYS}} and the standard deviation of the probability density function for {\sevensize \textit{PCIGALE}}.
  }
  \label{sl_figure6}
 \end{figure}
%%%%%%%%%%%%%%%%%%%%%%%%%%%%%%%%%%%
%  
 \citet{dacunha10b} investigated the dust--to--gas (D/G) mass ratio, M$_{D}$/ M$_{H}$, using the ratio of M$_{D}$ to SFR. The assumption is that the SFR can be used to infer the gas mass, using a Schmidt-Kennicutt law. We show the D/G mass ratio as a function of stellar mass in the upper panel of Fig.~\ref{sl_figure6}, using the SFR from the SED modelling and the Schmidt-Kennicutt law to obtain the gas mass. We have assumed the same star formation law as that used by \citet[][their eq.~10]{dacunha10b}, i.e. $\Sigma_{SFR}= (2.5\pm0.7)\times10^{-4}\Sigma_{H}^{1.4\pm0.15}$ \citep{Kennicutt98a}, where $\Sigma_{SFR}$ is the surface density of SFR, and $\Sigma_{H}$ is the surface density of the (HI+H$_{2}$) gas mass. We have applied the same factor as by \citet{dacunha10b} to correct from a \citet{Salpeter55} to a \citet{Chabrier03} initial mass function. The stellar mass can also be used to infer the metallicity, given their relation \citep[e.g.,][]{Tremonti04}, hence the D/G mass ratio as a function of stellar mass can indirectly provide information regarding the D/G mass ratio as a function of metallicity. This may be useful given the difficulty of obtaining gas-phase metallicities for a large sample of ETGs, including the current sample, while SED modelling can provide reliable information regarding stellar masses. Fig.~\ref{sl_figure6} shows that there is no clear correlation between stellar mass and D/G mass ratio in the case of ETGs; there is a higher scatter among the ETGs than that observed for the star-forming galaxies in Fig.~8 of \citet{dacunha10b}. This is the case for both results obtained with either {\sevensize \textit{MAGPHYS}} or {\sevensize \textit{PCIGALE}}. 

  Fig.~B.3 in \citet[][middle panels]{Remy14} shows the G/D mass ratio as a function of stellar mass for 126 local galaxies. Their dust masses are modelled as in \citet{Galliano11}, their gas masses are compiled from the literature, and their stellar masses are derived using the formula of \citet{Eskew12}. Their sample includes galaxies from the Dwarf Galaxy Survey \citep{Madden13,Remy13} and the KINGFISH survey \citep{Kennicutt11,Galametz11}. We derive a mean G/D mass ratio, $\langle$log$_{10}$(M$_{D,M}$/M$_{H})\rangle$\,=\,-1.1\,$\pm$\,0.5, and $\langle$log$_{10}$(M$_{D,P}$/M$_{H})\rangle$\,=\,-0.8\,$\pm$\,0.6, where the quoted uncertainties reflect the standard deviation. With a mean stellar mass of $\langle$log$_{10}$(M$_{\star,M})\rangle$\,=\,10.3\,$\pm$\,0.4\,M$_{\odot}$ and $\langle$log$_{10}$(M$_{\star,P})\rangle$\,=\,10.6\,$\pm$\,0.2\,M$_{\odot}$, the ETGs fall in the lower right part of Fig.~B.3 by \citet[][]{Remy14}, below their galaxies at that location, reflecting the lower gas reservoirs inferred from the SFR as compared to their dust masses.
 
 The lower panel of Fig.~\ref{sl_figure6} shows the observed-gas-to-modelled-dust mass ratio as a function of the stellar mass. In this panel, we choose to present the G/D mass ratio for a direct comparison with \citet[][middle panels]{Remy14}. The mean observed G/D mass ratio is $\langle$log$_{10}$(M$_{Gas}$/M$_{D,M})\rangle$\,=\,2.8\,$\pm$\,0.5, and  $\langle$log$_{10}$(M$_{Gas}$/M$_{D,P})\rangle$\,=\,2.4\,$\pm$\,0.6, now placing the ETGs at a location well within the galaxies studied in Fig.~B.3 by \citet[][]{Remy14}. The lower panel of Fig.~\ref{sl_figure6} shows a large scatter, as well. This large scatter is consistent with the idea that the ISM in ETGs is acquired from an external source. If the ISM is acquired from a satellite, then the G/D  mass ratio would indicate the properties of the ISM in the accreted satellite, showing no correlation with the properties of the ETG itself, as is the case in both panels in Fig.~\ref{sl_figure6}. Even though the scatter of the G/D mass ratio as a function of M$_{\star}$ seen in Fig.~\ref{sl_figure6} for the ETGs is large, nevertheless when they are placed in Fig.~B.3 of \citet[][]{Remy14} the trend of the G/D mass ratio decreasing with increasing M$_{\star}$ emerges, making the latter a useful tool to approximate the G/D mass ratio of ETGs knowing their M$_{\star}$. On the other hand, using a Schmidt-Kennicutt law similar to that describing disk galaxies together with an estimate of SFR through SED modelling, for example, underestimates the total gas mass and the G/D mass ratio of ETGs. We now explore the Schmidt-Kennicutt law for the ETGs.

%
%%%%% FIGURE 7 %%%%%%%%%% 
 \begin{figure}
  \centering
      \includegraphics[trim=3cm 2.5cm 3cm 3cm,width=8.5cm,clip]{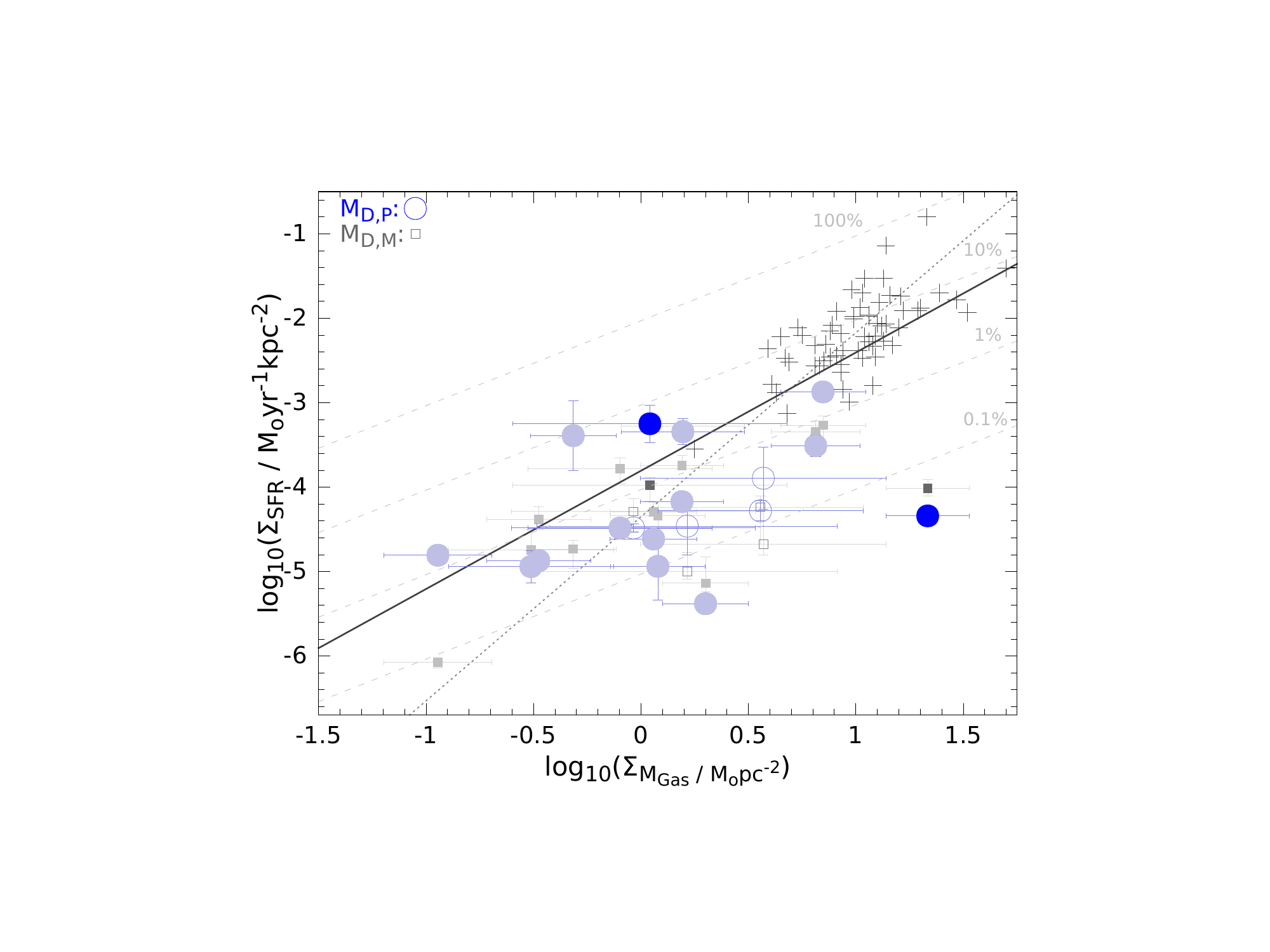}
      \caption{Surface density of star formation rate, $\sum_{SFR}$, versus surface density of the total gas mass, $\sum_{M_{Gas}}$. {\sevensize \textit{MAGPHYS}} results are shown in grey squares, while {\sevensize \textit{PCIGALE}} results in blue circles. Crosses correspond to galaxies taken from \citet{Kennicutt98a}. Dark-coloured filled symbols indicate ETGs with both \hbox{H\,{\sc i}} and CO; light-coloured filled symbols are ETGs with only \hbox{H\,{\sc i}}; open symbols are ETGs with only CO.
The dark--grey thick solid line is the Schmidt-Kennicutt law used here and by \citet{dacunha10b}. The light--grey thick solid line is the internal Milky Way star formation law from \citet{Misiriotis06}. The thin light-grey lines indicate the star formation efficiency per 10$^{8}$yr for values 0.1\%, 1\%, 10\%, and 100\%.
The grey error bars in all the data points represent the 16$^{th}$--84$^{th}$ percentile range for {\sevensize \textit{MAGPHYS}} and the standard deviation of the probability density function for  {\sevensize \textit{PCIGALE}}.
}
  \label{sl_figure7}
 \end{figure}
%%%%%%%%%%%%%%%%%%%%%%%%%%%%%%%%%%%
%
 In Fig.~\ref{sl_figure7} we show the surface density of the star formation rate, $\sum_{SFR}$ as a function of the surface density of the total gas mass, $\sum_{M_{Gas}}$, for the 18 galaxies in our sample. The thick dark--grey solid line is the Schmidt-Kennicutt law used here and by \citet{dacunha10b}; the thick light--grey solid line is the internal Milky Way star formation law from \citet{Misiriotis06}; The thin light-grey lines indicate the star formation efficiency per 10$^{8}$yr for values 0.1\%, 1\%, 10\%, and 100\%. \citet{Misiriotis06} obtained the Schmidt--Kennicutt law locally within the Milky Way and derived a power--law index equal to 2.18$\pm$0.20 (see eq. 16 of \citealt{Misiriotis06}). \citet{Misiriotis06} called this star formation law for the Milky Way an internal Schmidt law.

 The ETGs of our sample extend towards lower gas masses and SFRs and show a larger scatter as compared to the internal Milky Way star formation law of \citet{Misiriotis06}, or the galaxies studied by \citet{Kennicutt98a} shown in Fig.~\ref{sl_figure7}. The majority of the ETGs have star formation efficiency per 0.1\,Gyr between 1\% to 0.1\%, i.e. lower than that of normal galaxies. In addition, the ETGs are offset from the Schmidt-Kennicutt law, independent of the SED model used to derive SFRs. \citet{Martig13} investigate the Schmidt-Kennicutt law via simulations of gas disks embedded in ETGs, and reach similar conclusions regarding the star formation efficiency and scatter, signifying the process of morphological quenching \citep{Martig09}. The high scatter and the deviation from both the internal Milky Way star formation law and the Schmidt-Kennicutt law followed by disk galaxies indicate the complex star formation and ISM histories of the ETGs. Indeed, the majority of the galaxies studied here show many morphological signs of tidal interactions in FIR/submm imaging.

   Past studies investigating the star formation law of ETGs include the work of \citet{Combes07}, \citet{Crocker11}, and \citet{Davis14}. The latter authors study molecular gas-rich ETGs, and investigate their star formation law using their total gas mass and SFR tracer based on FUV+22$\mu$m. These authors find overall higher SFR surface densities for their ETGs, comparable to those of normal spiral galaxies (see their Fig.~5), but they lie systematically lower than a Schmidt-Kennicutt law, indicating lower star formation efficiency. Lower star formation efficiency is also found for molecular gas-rich dust-lane ETGs by \citet{Davis15}, who suggest that the lower star formation efficiency is due to a suppression of star formation by minor mergers. They assume that the gas and dust in these dust-lane ETGs comes from the accreted galaxy through minor merging, and they use the observed gas--to--dust mass ratio as a function of stellar mass to infer the stellar mass of the accreted galaxy \citep[see Fig.~3 in][]{Davis15}.

  For the dustier galaxies studied here, given the scatter of the cold dust--rich ETGs when placing them in the Schmidt-Kennicutt law, their SFR is not a good tracer of their gas mass. The SFR underestimates ETGs' gas mass in the case of a power--law index appropriate for disk galaxies, as used by \citet{dacunha10b} and here for the upper panel of Fig.~\ref{sl_figure6}. 
%%%%% FIGURE 8   %%%%%%%%%% 
 \begin{figure}
  \centering
      \includegraphics[trim=3cm 2.7cm 3cm 3cm,width=8.5cm,clip]{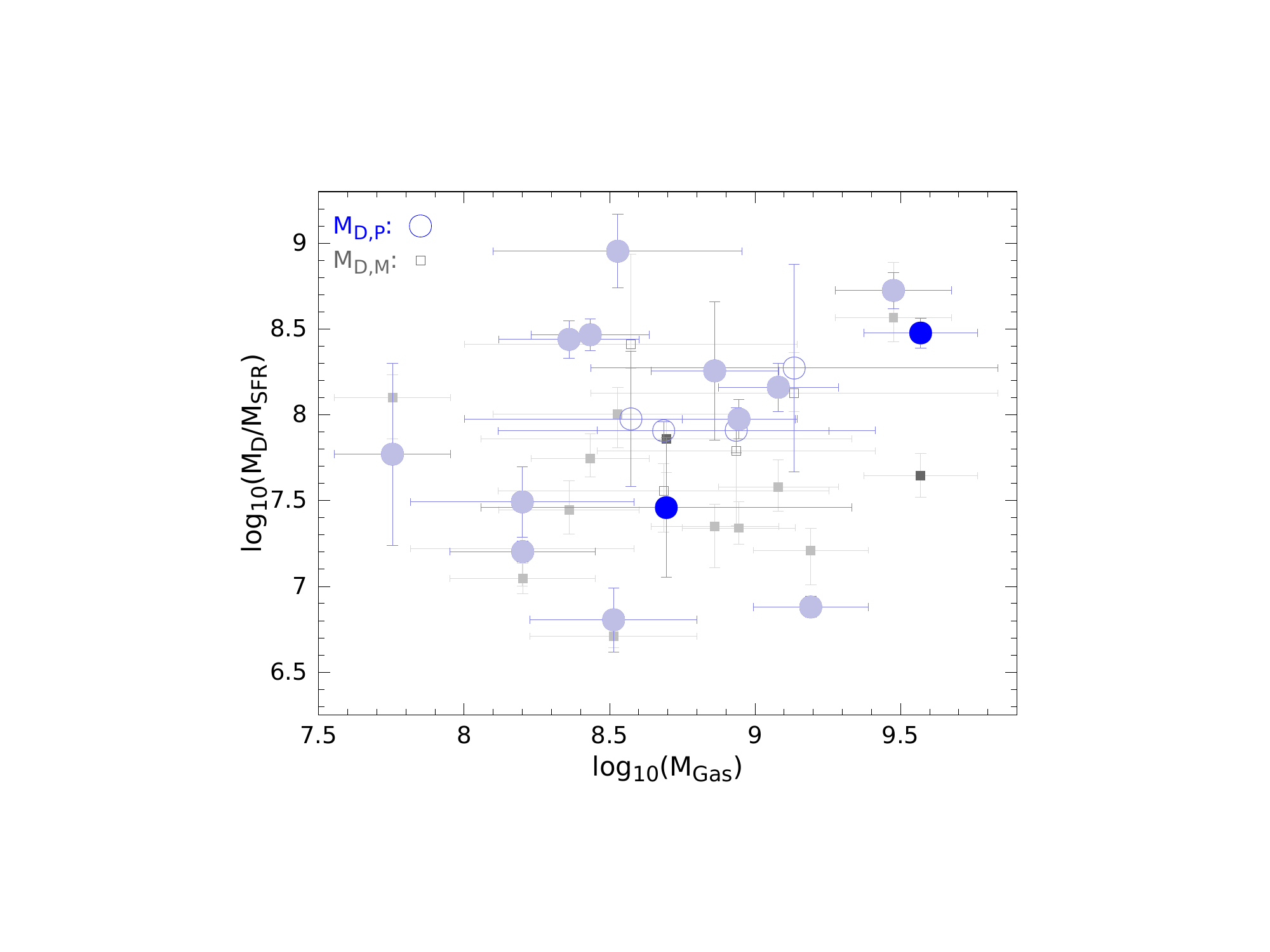}
      \includegraphics[trim=3cm 2.7cm 3cm 3cm,width=8.5cm,clip]{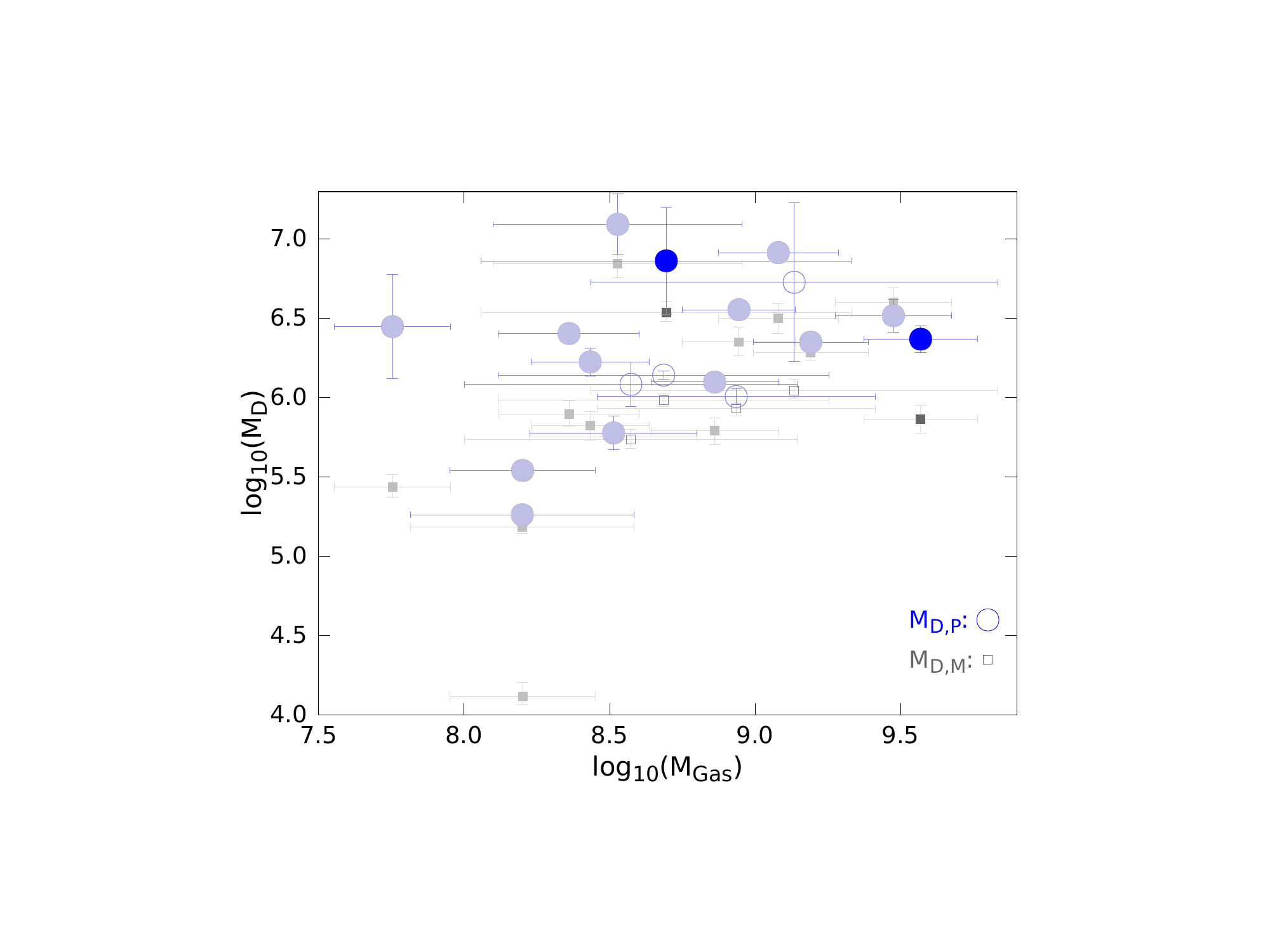}
      \caption{\textit{ Upper panel}: M$_{D}$/SFR as a function of the total gas mass.
\textit{ Lower panel}:  M$_{D}$ as a function of the total gas mass.
    In all panels, {\sevensize \textit{MAGPHYS}} results are shown in grey squares, while {\sevensize \textit{PCIGALE}} results in blue circles. Dark-coloured filled symbols indicate ETGs with both \hbox{H\,{\sc i}} and CO; light-coloured filled symbols are ETGs with only \hbox{H\,{\sc i}}; open symbols are ETGs with only CO.
    The grey error bars represent the 16$^{th}$--84$^{th}$ percentile range for {\sevensize \textit{MAGPHYS}} or the standard deviation for {\sevensize \textit{PCIGALE}}.
  }
  \label{sl_figure8}
 \end{figure}
%%%%%%%%%%%%%%%%%%%%%%%%%%%%%%%%%%%
 %
 Indeed, the upper panel of Fig.~\ref{sl_figure8} shows that there is no correlation between the ratio M$_{D}$/SFR and the total gas mass, M$_{Gas}$. For a given value of the total gas mass, there is between 1 to 3\,dex difference in the ratio M$_{D}$/SFR. This latter ratio, as a property derived from the SED modelling, may serve as a practical proxy to infer indirectly the D/G mass ratio, as discussed earlier (by assuming that the gas mass is derived through the SFR assuming a Schmidt--Kennicutt law). In the case of ETGs, though, this ratio provides a poor proxy to estimate their D/G mass ratio.

 In the lower panel of Fig.~\ref{sl_figure8} we explore how the observed total gas mass M$_{Gas}$ relates to the dust mass, M$_{D}$. This figure reveals that there is no trend between the dust mass and the gas mass, as for a given M$_{Gas}$ there is between 1 to 4\,dex scatter in M$_{D}$. These results suggest that assuming a dust mass for ETGs will not provide a good estimate of their M$_{Gas}$.
%______________________________________________________________

\section{Summary \& conclusions}
\label{sec:summary}

We analyze UV--to--submm imaging for a sample of 18 early-type galaxies, drawn from the Herschel Reference Survey \citep{Boselli10}, in order to derive their star formation and dust properties via SED modelling. The sample is selected to be detected both in FIR/submm with all three \textit{ Herschel}/SPIRE bands, and in either \hbox{H\,{\sc i}} or \hbox{CO} or both. We combine the dust and star formation results with literature information on their atomic and molecular gas masses \citep{Boselli14a}, in order to relate the dust and star formation properties with the total gas mass in these ETGs. Our major findings are summarized as follows: 
   \begin{enumerate}
   \item When the ETGs are compared to the M$_{D}$--SFR relation for disk galaxies studied by \citet{dacunha10b}, they do not follow the disk galaxies' correlation and show a large scatter beyond the intrinsic scatter of the relation (Fig.~\ref{sl_figure5}).
   \item When placing the ETGs in the stellar mass versus SFR plane, they fall below the relation defined by normal star forming galaxies, showing much smaller specific SFRs, by a factor between 30 to 4000 smaller than that of the normal star-forming galaxies (Fig.~\ref{sl_figure5b}). 
   \item The ETGs studied here do not follow the same Schmidt--Kennicutt law as the disk galaxies studied by \citet{dacunha10b}, neither do they follow the internal Schmidt law found locally within Milky Way by \citet{Misiriotis06}, showing a large scatter and extending to lower SFRs and gas masses (Fig.~\ref{sl_figure7}).
   \item The Schmidt--Kennicutt law valid for disk galaxies \citep{dacunha10b,Kennicutt98a} can not be used along with the SFR to infer the total gas mass of ETGs, as the total gas masses inferred are under-predicted (upper panel of Fig.~\ref{sl_figure6}). Contrary to the case of disk galaxies, this renders the ratio M$_{D}$/SFR  a poor proxy for inferring the D/G mass ratio of ETGs (Fig.~\ref{sl_figure8}).
   \item The total gas masses derived observationally (i.e., using eq.~2) provide G/D mass ratios consistent, on average, with the ones found for the galaxy sample studied by \citet{Remy14}, and fall in the right place for their M$_{\star}$, on average, as those galaxies studied by \citet[][their figure B.3]{Remy14}, making the latter a useful relation to approximate the G/D mass ratio of ETGs (lower panel of Fig.~\ref{sl_figure6}). The large scatter seen in individual ETGs in Fig.~\ref{sl_figure6}, though, indicates an externally acquired origin for their ISM.
   \end{enumerate}
   
  ETGs have complex ISM and star formation histories. This complexity is a result of their merging and satellite accretion history. The majority of the ETGs studied here show morphology of recent interactions. This is not uncommon among ETGs \citep[e.g.,][]{Dariush16}, and their ISM history coupled to the satellite accretion history has been uncovered with deep optical imaging from the ATLAS$^{3D}$ survey \citep{Duc15}, as well as imprinted in their colour properties \citep{Young14}. Such interaction events may be the cause of star formation responsible for significant dust production \citep{Wilson13}, hence their star formation history coupled to their ISM history leads to deviation or significant scatter from a Schmidt--Kennicutt law relation.  
%______________________________________________________________

\section*{Acknowledgements}

We are grateful to an anonymous referee for useful suggestions, which helped us to improve the manuscript. Support for the work of SL and PB is provided by an NSERC Discovery Grant and by the Academic Development Fund of the University of Western Ontario. SL thanks M\'{e}d\'{e}ric Boquien for support with {\sevensize \textit{PCIGALE}}, and Pierre-Alain Duc for useful discussions. SL, EX, and SM acknowledge support from DustPedia, a collaborative focused research project supported by the European Union under the Seventh Framework Programme (2007-2013) call (proposal no. 606847), with participating institutions: Cardiff University, UK; National Observatory of Athens, Greece; Ghent University, Belgium; Universit\'{e} Paris Sud, France; National Institute for Astrophysics, Italy and CEA (Paris), France.

   This research has made use of data from HRS project. HRS is a Herschel Key Programme utilizing Guaranteed Time from the SPIRE instrument team, ESAC scientists and a mission scientist.
The HRS data was accessed through the Herschel Database in Marseilles (HeDaM; \url{http://hedam.lam.fr}) operated by CeSAM and hosted by the Laboratoire d'Astrophysique de Marseilles.

   This research made use of several facilities and open source software: Astropy \citep[\url{http://www.astropy.org}]{Astropy13}, a community-developed core Python package for Astronomy; APLpy (\url{http://aplpy.github.com}), an open-source plotting package for Python; IRAF, distributed by the National Optical Astronomy Observatory, which is operated by the Association of Universities for Research in Astronomy (AURA) under cooperative agreement with the National Science Foundation; STSDAS and PyRAF, products of the Space Telescope Science Institute, which is operated by AURA for NASA; NASA/IPAC Extragalactic Database (NED), operated by the Jet Propulsion Laboratory, California Institute of Technology, under contract with the National Aeronautics and Space Administration; NASA/IPAC Infrared Science Archive (IRSA), operated by the Jet Propulsion Laboratory, California Institute of Technology, under contract with the National Aeronautics and Space Administration; Aladin; NASA's Astrophysics Data System Bibliographic Services; SAOImage DS9, developed by Smithsonian Astrophysical Observatory; the Spanish Virtual Observatory (\url{http://svo.cab.inta-csic.es}) supported from the Spanish MICINN / MINECO through grants AyA2008-02156, AyA2011-24052. This work made extensive use of the free software GNU Octave and the authors are grateful to the Octave development community for their support. 
    
\bibliographystyle{mnras}
\bibliography{160615ETGsSR}{}

\begin{thebibliography}{}
\makeatletter
\relax
\def\mn@urlcharsother{\let\do\@makeother \do\$\do\&\do\#\do\^\do\_\do\%\do\~}
\def\mn@doi{\begingroup\mn@urlcharsother \@ifnextchar [ {\mn@doi@}
  {\mn@doi@[]}}
\def\mn@doi@[#1]#2{\def\@tempa{#1}\ifx\@tempa\@empty \href
  {http://dx.doi.org/#2} {doi:#2}\else \href {http://dx.doi.org/#2} {#1}\fi
  \endgroup}
\def\mn@eprint#1#2{\mn@eprint@#1:#2::\@nil}
\def\mn@eprint@arXiv#1{\href {http://arxiv.org/abs/#1} {{\tt arXiv:#1}}}
\def\mn@eprint@dblp#1{\href {http://dblp.uni-trier.de/rec/bibtex/#1.xml}
  {dblp:#1}}
\def\mn@eprint@#1:#2:#3:#4\@nil{\def\@tempa {#1}\def\@tempb {#2}\def\@tempc
  {#3}\ifx \@tempc \@empty \let \@tempc \@tempb \let \@tempb \@tempa \fi \ifx
  \@tempb \@empty \def\@tempb {arXiv}\fi \@ifundefined
  {mn@eprint@\@tempb}{\@tempb:\@tempc}{\expandafter \expandafter \csname
  mn@eprint@\@tempb\endcsname \expandafter{\@tempc}}}

\bibitem[\protect\citeauthoryear{{Aniano}, {Draine}, {Gordon}  \&
  {Sandstrom}}{{Aniano} et~al.}{2011}]{Aniano11}
{Aniano} G.,  {Draine} B.~T.,  {Gordon} K.~D.,   {Sandstrom} K.,  2011, \mn@doi
  [\pasp] {10.1086/662219}, \href
  {http://cdsads.u-strasbg.fr/abs/2011PASP..123.1218A} {123, 1218}

\bibitem[\protect\citeauthoryear{{Aniano}, {Draine}, {Calzetti}  \& {et
  al.}}{{Aniano} et~al.}{2012}]{Aniano12}
{Aniano} G.,  {Draine} B.~T.,  {Calzetti} D.,   {et al.} 2012, \mn@doi [\apj]
  {10.1088/0004-637X/756/2/138}, \href
  {http://adsabs.harvard.edu/abs/2012ApJ...756..138A} {756, 138}

\bibitem[\protect\citeauthoryear{{Annibali}, {Bressan}, {Rampazzo},
  {Zeilinger}, {Vega}  \& {Panuzzo}}{{Annibali} et~al.}{2010}]{Annibali10}
{Annibali} F.,  {Bressan} A.,  {Rampazzo} R.,  {Zeilinger} W.~W.,  {Vega} O.,
  {Panuzzo} P.,  2010, \mn@doi [\aap] {10.1051/0004-6361/200913774}, \href
  {http://adsabs.harvard.edu/abs/2010A%26A...519A..40A} {519, A40}

\bibitem[\protect\citeauthoryear{{Appleton} et~al.,}{{Appleton}
  et~al.}{2014}]{Appleton14}
{Appleton} P.~N.,  et~al., 2014, \mn@doi [\apj] {10.1088/0004-637X/797/2/117},
  \href {http://cdsads.u-strasbg.fr/abs/2014ApJ...797..117A} {797, 117}

\bibitem[\protect\citeauthoryear{{Astropy Collaboration}, {Robitaille},
  {Tollerud}  \& {et al.}}{{Astropy Collaboration} et~al.}{2013}]{Astropy13}
{Astropy Collaboration} {Robitaille} T.~P.,  {Tollerud} E.~J.,   {et al.} 2013,
  \mn@doi [\aap] {10.1051/0004-6361/201322068}, \href
  {http://adsabs.harvard.edu/abs/2013A%26A...558A..33A} {558, A33}

\bibitem[\protect\citeauthoryear{{Bendo}, {Galliano}  \& {Madden}}{{Bendo}
  et~al.}{2012}]{Bendo12}
{Bendo} G.~J.,  {Galliano} F.,   {Madden} S.~C.,  2012, \mn@doi [\mnras]
  {10.1111/j.1365-2966.2012.20784.x}, \href
  {http://cdsads.u-strasbg.fr/abs/2012MNRAS.423..197B} {423, 197}

\bibitem[\protect\citeauthoryear{{Berta}, {Lutz}, {Genzel},
  {F{\"o}rster-Schreiber}  \& {Tacconi}}{{Berta} et~al.}{2016}]{Berta16}
{Berta} S.,  {Lutz} D.,  {Genzel} R.,  {F{\"o}rster-Schreiber} N.~M.,
  {Tacconi} L.~J.,  2016, \mn@doi [\aap] {10.1051/0004-6361/201527746}, \href
  {http://cdsads.u-strasbg.fr/abs/2016A%26A...587A..73B} {587, A73}

\bibitem[\protect\citeauthoryear{{Bianchi}}{{Bianchi}}{2013}]{Bianchi13}
{Bianchi} S.,  2013, \mn@doi [\aap] {10.1051/0004-6361/201220866}, \href
  {http://cdsads.u-strasbg.fr/abs/2013A%26A...552A..89B} {552, A89}

\bibitem[\protect\citeauthoryear{{Boquien}, {Buat}, {Boselli}  \& {et
  al.}}{{Boquien} et~al.}{2012}]{Boquien12}
{Boquien} M.,  {Buat} V.,  {Boselli} A.,   {et al.} 2012, \mn@doi [\aap]
  {10.1051/0004-6361/201118624}, \href
  {http://cdsads.u-strasbg.fr/abs/2012A%26A...539A.145B} {539, A145}

\bibitem[\protect\citeauthoryear{{Boquien}, {Boselli}, {Buat}  \& {et
  al.}}{{Boquien} et~al.}{2013}]{Boquien13}
{Boquien} M.,  {Boselli} A.,  {Buat} V.,   {et al.} 2013, \mn@doi [\aap]
  {10.1051/0004-6361/201220768}, \href
  {http://adsabs.harvard.edu/abs/2013A%26A...554A..14B} {554, A14}

\bibitem[\protect\citeauthoryear{{Boquien} et~al.,}{{Boquien}
  et~al.}{2016}]{Boquien16}
{Boquien} M.,  et~al., 2016, \mn@doi [\aap] {10.1051/0004-6361/201527759},
  \href {http://cdsads.u-strasbg.fr/abs/2016A%26A...591A...6B} {591, A6}

\bibitem[\protect\citeauthoryear{{Boselli}, {Lequeux}  \& {Gavazzi}}{{Boselli}
  et~al.}{2002}]{Boselli02}
{Boselli} A.,  {Lequeux} J.,   {Gavazzi} G.,  2002, \mn@doi [\aap]
  {10.1051/0004-6361:20011747}, \href
  {http://adsabs.harvard.edu/abs/2002A%26A...384...33B} {384, 33}

\bibitem[\protect\citeauthoryear{{Boselli}, {Eales}, {Cortese}  \& {et
  al.}}{{Boselli} et~al.}{2010}]{Boselli10}
{Boselli} A.,  {Eales} S.,  {Cortese} L.,   {et al.} 2010, \mn@doi [\pasp]
  {10.1086/651535}, \href {http://adsabs.harvard.edu/abs/2010PASP..122..261B}
  {122, 261}

\bibitem[\protect\citeauthoryear{{Boselli}, {Cortese}  \& {Boquien}}{{Boselli}
  et~al.}{2014a}]{Boselli14a}
{Boselli} A.,  {Cortese} L.,   {Boquien} M.,  2014a, \mn@doi [\aap]
  {10.1051/0004-6361/201322311}, \href
  {http://adsabs.harvard.edu/abs/2014A%26A...564A..65B} {564, A65}

\bibitem[\protect\citeauthoryear{{Boselli}, {Cortese}, {Boquien}  \& {et
  al.}}{{Boselli} et~al.}{2014b}]{Boselli14b}
{Boselli} A.,  {Cortese} L.,  {Boquien} M.,   {et al.} 2014b, \mn@doi [\aap]
  {10.1051/0004-6361/201322312}, \href
  {http://cdsads.u-strasbg.fr/abs/2014A%26A...564A..66B} {564, A66}

\bibitem[\protect\citeauthoryear{{Boselli}, {Cortese}, {Boquien}  \& {et
  al.}}{{Boselli} et~al.}{2014c}]{Boselli14c}
{Boselli} A.,  {Cortese} L.,  {Boquien} M.,   {et al.} 2014c, \mn@doi [\aap]
  {10.1051/0004-6361/201322313}, \href
  {http://cdsads.u-strasbg.fr/abs/2014A%26A...564A..67B} {564, A67}

\bibitem[\protect\citeauthoryear{{Boselli}, {Fossati}, {Gavazzi}  \& {et
  al.}}{{Boselli} et~al.}{2015}]{Boselli15a}
{Boselli} A.,  {Fossati} M.,  {Gavazzi} G.,   {et al.} 2015, \mn@doi [\aap]
  {10.1051/0004-6361/201525712}, \href
  {http://cdsads.u-strasbg.fr/abs/2015A%26A...579A.102B} {579, A102}

\bibitem[\protect\citeauthoryear{{Bresolin}}{{Bresolin}}{2013}]{Bresolin13}
{Bresolin} F.,  2013, \mn@doi [\apjl] {10.1088/2041-8205/772/2/L23}, \href
  {http://adsabs.harvard.edu/abs/2013ApJ...772L..23B} {772, L23}

\bibitem[\protect\citeauthoryear{{Bressan} et~al.,}{{Bressan}
  et~al.}{2006}]{Bressan06}
{Bressan} A.,  et~al., 2006, \mn@doi [\apjl] {10.1086/502970}, \href
  {http://cdsads.u-strasbg.fr/abs/2006ApJ...639L..55B} {639, L55}

\bibitem[\protect\citeauthoryear{{Brinchmann}, {Charlot}, {White}  \& {et
  al.}}{{Brinchmann} et~al.}{2004}]{Brinchmann04}
{Brinchmann} J.,  {Charlot} S.,  {White} S.~D.~M.,   {et al.} 2004, \mn@doi
  [\mnras] {10.1111/j.1365-2966.2004.07881.x}, \href
  {http://cdsads.u-strasbg.fr/abs/2004MNRAS.351.1151B} {351, 1151}

\bibitem[\protect\citeauthoryear{{Bruzual} \& {Charlot}}{{Bruzual} \&
  {Charlot}}{2003}]{Bruzual03}
{Bruzual} G.,  {Charlot} S.,  2003, \mn@doi [\mnras]
  {10.1046/j.1365-8711.2003.06897.x}, \href
  {http://adsabs.harvard.edu/abs/2003MNRAS.344.1000B} {344, 1000}

\bibitem[\protect\citeauthoryear{{Burgarella}, {Buat}  \&
  {Iglesias-P{\'a}ramo}}{{Burgarella} et~al.}{2005}]{Burgarella05}
{Burgarella} D.,  {Buat} V.,   {Iglesias-P{\'a}ramo} J.,  2005, \mn@doi
  [\mnras] {10.1111/j.1365-2966.2005.09131.x}, \href
  {http://cdsads.u-strasbg.fr/abs/2005MNRAS.360.1413B} {360, 1413}

\bibitem[\protect\citeauthoryear{{Calzetti}, {Armus}, {Bohlin}, {Kinney},
  {Koornneef}  \& {Storchi-Bergmann}}{{Calzetti} et~al.}{2000}]{Calzetti00}
{Calzetti} D.,  {Armus} L.,  {Bohlin} R.~C.,  {Kinney} A.~L.,  {Koornneef} J.,
   {Storchi-Bergmann} T.,  2000, \mn@doi [\apj] {10.1086/308692}, \href
  {http://cdsads.u-strasbg.fr/abs/2000ApJ...533..682C} {533, 682}

\bibitem[\protect\citeauthoryear{{Carretta} \& {Gratton}}{{Carretta} \&
  {Gratton}}{1997}]{Carretta97}
{Carretta} E.,  {Gratton} R.~G.,  1997, \mn@doi [\aaps] {10.1051/aas:1997116},
  \href {http://cdsads.u-strasbg.fr/abs/1997A%26AS..121...95C} {121, 95}

\bibitem[\protect\citeauthoryear{{Chabrier}}{{Chabrier}}{2003}]{Chabrier03}
{Chabrier} G.,  2003, \mn@doi [\pasp] {10.1086/376392}, \href
  {http://cdsads.u-strasbg.fr/abs/2003PASP..115..763C} {115, 763}

\bibitem[\protect\citeauthoryear{{Charlot} \& {Fall}}{{Charlot} \&
  {Fall}}{2000}]{Charlot00}
{Charlot} S.,  {Fall} S.~M.,  2000, \mn@doi [\apj] {10.1086/309250}, \href
  {http://adsabs.harvard.edu/abs/2000ApJ...539..718C} {539, 718}

\bibitem[\protect\citeauthoryear{{Charmandaris}, {Combes}  \& {van der
  Hulst}}{{Charmandaris} et~al.}{2000}]{Charmandaris00}
{Charmandaris} V.,  {Combes} F.,   {van der Hulst} J.~M.,  2000, \aap, \href
  {http://adsabs.harvard.edu/abs/2000A%26A...356L...1C} {356, L1}

\bibitem[\protect\citeauthoryear{{Ciesla}, {Boselli}, {Smith}  \& {et
  al.}}{{Ciesla} et~al.}{2012}]{Ciesla12}
{Ciesla} L.,  {Boselli} A.,  {Smith} M.~W.~L.,   {et al.} 2012, \mn@doi [\aap]
  {10.1051/0004-6361/201219216}, \href
  {http://adsabs.harvard.edu/abs/2012A%26A...543A.161C} {543, A161}

\bibitem[\protect\citeauthoryear{{Ciesla}, {Boquien}, {Boselli}  \& {et
  al.}}{{Ciesla} et~al.}{2014}]{Ciesla14}
{Ciesla} L.,  {Boquien} M.,  {Boselli} A.,   {et al.} 2014, \mn@doi [\aap]
  {10.1051/0004-6361/201323248}, \href
  {http://adsabs.harvard.edu/abs/2014A%26A...565A.128C} {565, A128}

\bibitem[\protect\citeauthoryear{{Ciesla}, {Charmandaris}, {Georgakakis}  \&
  {et al.}}{{Ciesla} et~al.}{2015}]{Ciesla15}
{Ciesla} L.,  {Charmandaris} V.,  {Georgakakis} A.,   {et al.} 2015, \mn@doi
  [\aap] {10.1051/0004-6361/201425252}, \href
  {http://cdsads.u-strasbg.fr/abs/2015A%26A...576A..10C} {576, A10}

\bibitem[\protect\citeauthoryear{{Clemens}, {Jones}, {Bressan}  \& {et
  al.}}{{Clemens} et~al.}{2010}]{Clemens10}
{Clemens} M.~S.,  {Jones} A.~P.,  {Bressan} A.,   {et al.} 2010, \mn@doi [\aap]
  {10.1051/0004-6361/201014533}, \href
  {http://cdsads.u-strasbg.fr/abs/2010A%26A...518L..50C} {518, L50}

\bibitem[\protect\citeauthoryear{{Combes}, {Young}  \& {Bureau}}{{Combes}
  et~al.}{2007}]{Combes07}
{Combes} F.,  {Young} L.~M.,   {Bureau} M.,  2007, \mn@doi [\mnras]
  {10.1111/j.1365-2966.2007.11759.x}, \href
  {http://adsabs.harvard.edu/abs/2007MNRAS.377.1795C} {377, 1795}

\bibitem[\protect\citeauthoryear{{Cortese}, {Ciesla}, {Boselli}  \& {et
  al.}}{{Cortese} et~al.}{2012a}]{Cortese12b}
{Cortese} L.,  {Ciesla} L.,  {Boselli} A.,   {et al.} 2012a, \mn@doi [\aap]
  {10.1051/0004-6361/201118499}, \href
  {http://adsabs.harvard.edu/abs/2012A%26A...540A..52C} {540, A52}

\bibitem[\protect\citeauthoryear{{Cortese}, {Boissier}, {Boselli}  \& {et
  al.}}{{Cortese} et~al.}{2012b}]{Cortese12a}
{Cortese} L.,  {Boissier} S.,  {Boselli} A.,   {et al.} 2012b, \mn@doi [\aap]
  {10.1051/0004-6361/201219312}, \href
  {http://adsabs.harvard.edu/abs/2012A%26A...544A.101C} {544, A101}

\bibitem[\protect\citeauthoryear{{Cortese}, {Fritz}, {Bianchi}  \& {et
  al.}}{{Cortese} et~al.}{2014}]{Cortese14}
{Cortese} L.,  {Fritz} J.,  {Bianchi} S.,   {et al.} 2014, \mn@doi [\mnras]
  {10.1093/mnras/stu175}, \href
  {http://adsabs.harvard.edu/abs/2014MNRAS.440..942C} {440, 942}

\bibitem[\protect\citeauthoryear{{Crocker}, {Bureau}, {Young}  \&
  {Combes}}{{Crocker} et~al.}{2011}]{Crocker11}
{Crocker} A.~F.,  {Bureau} M.,  {Young} L.~M.,   {Combes} F.,  2011, \mn@doi
  [\mnras] {10.1111/j.1365-2966.2010.17537.x}, \href
  {http://adsabs.harvard.edu/abs/2011MNRAS.410.1197C} {410, 1197}

\bibitem[\protect\citeauthoryear{{Dale} et~al.,}{{Dale} et~al.}{2012}]{Dale12}
{Dale} D.~A.,  et~al., 2012, \mn@doi [\apj] {10.1088/0004-637X/745/1/95}, \href
  {http://cdsads.u-strasbg.fr/abs/2012ApJ...745...95D} {745, 95}

\bibitem[\protect\citeauthoryear{{Dariush} et~al.,}{{Dariush}
  et~al.}{2016}]{Dariush16}
{Dariush} A.,  et~al., 2016, \mn@doi [\mnras] {10.1093/mnras/stv2767}, \href
  {http://cdsads.u-strasbg.fr/abs/2016MNRAS.456.2221D} {456, 2221}

\bibitem[\protect\citeauthoryear{{Davies}, {Baes}, {Bendo}  \& {et
  al.}}{{Davies} et~al.}{2010}]{Davies10}
{Davies} J.~I.,  {Baes} M.,  {Bendo} G.~J.,   {et al.} 2010, \mn@doi [\aap]
  {10.1051/0004-6361/201014571}, \href
  {http://cdsads.u-strasbg.fr/abs/2010A%26A...518L..48D} {518, L48}

\bibitem[\protect\citeauthoryear{{Davis}, {Alatalo}, {Sarzi}  \& {et
  al.}}{{Davis} et~al.}{2011}]{Davis11}
{Davis} T.~A.,  {Alatalo} K.,  {Sarzi} M.,   {et al.} 2011, \mn@doi [\mnras]
  {10.1111/j.1365-2966.2011.19355.x}, \href
  {http://adsabs.harvard.edu/abs/2011MNRAS.417..882D} {417, 882}

\bibitem[\protect\citeauthoryear{{Davis}, {Young}, {Crocker}  \& {et
  al.}}{{Davis} et~al.}{2014}]{Davis14}
{Davis} T.~A.,  {Young} L.~M.,  {Crocker} A.~F.,   {et al.} 2014, \mn@doi
  [\mnras] {10.1093/mnras/stu570}, \href
  {http://adsabs.harvard.edu/abs/2014MNRAS.444.3427D} {444, 3427}

\bibitem[\protect\citeauthoryear{{Davis}, {Rowlands}, {Allison}  \& {et
  al.}}{{Davis} et~al.}{2015}]{Davis15}
{Davis} T.~A.,  {Rowlands} K.,  {Allison} J.~R.,   {et al.} 2015, \mn@doi
  [\mnras] {10.1093/mnras/stv597}, \href
  {http://adsabs.harvard.edu/abs/2015MNRAS.449.3503D} {449, 3503}

\bibitem[\protect\citeauthoryear{{Draine} \& {Li}}{{Draine} \&
  {Li}}{2007}]{Draine07}
{Draine} B.~T.,  {Li} A.,  2007, \mn@doi [\apj] {10.1086/511055}, \href
  {http://cdsads.u-strasbg.fr/abs/2007ApJ...657..810D} {657, 810}

\bibitem[\protect\citeauthoryear{{Duc}, {Cuillandre}, {Karabal}  \& {et
  al.}}{{Duc} et~al.}{2015}]{Duc15}
{Duc} P.-A.,  {Cuillandre} J.-C.,  {Karabal} E.,   {et al.} 2015, \mn@doi
  [\mnras] {10.1093/mnras/stu2019}, \href
  {http://cdsads.u-strasbg.fr/abs/2015MNRAS.446..120D} {446, 120}

\bibitem[\protect\citeauthoryear{{Elbaz}, {Daddi}, {Le Borgne}  \& {et
  al.}}{{Elbaz} et~al.}{2007}]{Elbaz07}
{Elbaz} D.,  {Daddi} E.,  {Le Borgne} D.,   {et al.} 2007, \mn@doi [\aap]
  {10.1051/0004-6361:20077525}, \href
  {http://cdsads.u-strasbg.fr/abs/2007A%26A...468...33E} {468, 33}

\bibitem[\protect\citeauthoryear{{Elbaz}, {Dickinson}, {Hwang}  \& {et
  al.}}{{Elbaz} et~al.}{2011}]{Elbaz11}
{Elbaz} D.,  {Dickinson} M.,  {Hwang} H.~S.,   {et al.} 2011, \mn@doi [\aap]
  {10.1051/0004-6361/201117239}, \href
  {http://cdsads.u-strasbg.fr/abs/2011A%26A...533A.119E} {533, A119}

\bibitem[\protect\citeauthoryear{{Eskew}, {Zaritsky}  \& {Meidt}}{{Eskew}
  et~al.}{2012}]{Eskew12}
{Eskew} M.,  {Zaritsky} D.,   {Meidt} S.,  2012, \mn@doi [\aj]
  {10.1088/0004-6256/143/6/139}, \href
  {http://cdsads.u-strasbg.fr/abs/2012AJ....143..139E} {143, 139}

\bibitem[\protect\citeauthoryear{{Fazio}, {Hora}, {Allen}  \& {et al.}}{{Fazio}
  et~al.}{2004}]{Fazio04}
{Fazio} G.~G.,  {Hora} J.~L.,  {Allen} L.~E.,   {et al.} 2004, \mn@doi [\apjs]
  {10.1086/422843}, \href {http://adsabs.harvard.edu/abs/2004ApJS..154...10F}
  {154, 10}

\bibitem[\protect\citeauthoryear{{Feltre}, {Hatziminaoglou}, {Fritz}  \&
  {Franceschini}}{{Feltre} et~al.}{2012}]{Feltre12}
{Feltre} A.,  {Hatziminaoglou} E.,  {Fritz} J.,   {Franceschini} A.,  2012,
  \mn@doi [\mnras] {10.1111/j.1365-2966.2012.21695.x}, \href
  {http://cdsads.u-strasbg.fr/abs/2012MNRAS.426..120F} {426, 120}

\bibitem[\protect\citeauthoryear{{Fitzpatrick}}{{Fitzpatrick}}{1999}]{Fitzpatrick99}
{Fitzpatrick} E.~L.,  1999, \mn@doi [\pasp] {10.1086/316293}, \href
  {http://adsabs.harvard.edu/abs/1999PASP..111...63F} {111, 63}

\bibitem[\protect\citeauthoryear{{Forbes}, {Spitler}, {Strader}  \& {et
  al.}}{{Forbes} et~al.}{2011}]{Forbes11}
{Forbes} D.~A.,  {Spitler} L.~R.,  {Strader} J.,   {et al.} 2011, \mn@doi
  [\mnras] {10.1111/j.1365-2966.2011.18373.x}, \href
  {http://cdsads.u-strasbg.fr/abs/2011MNRAS.413.2943F} {413, 2943}

\bibitem[\protect\citeauthoryear{{Ford} \& {Bregman}}{{Ford} \&
  {Bregman}}{2013}]{Ford13}
{Ford} H.~A.,  {Bregman} J.~N.,  2013, \mn@doi [\apj]
  {10.1088/0004-637X/770/2/137}, \href
  {http://adsabs.harvard.edu/abs/2013ApJ...770..137F} {770, 137}

\bibitem[\protect\citeauthoryear{{Fritz}, {Franceschini}  \&
  {Hatziminaoglou}}{{Fritz} et~al.}{2006}]{Fritz06}
{Fritz} J.,  {Franceschini} A.,   {Hatziminaoglou} E.,  2006, \mn@doi [\mnras]
  {10.1111/j.1365-2966.2006.09866.x}, \href
  {http://cdsads.u-strasbg.fr/abs/2006MNRAS.366..767F} {366, 767}

\bibitem[\protect\citeauthoryear{{Galametz}, {Madden}, {Galliano}  \& {et
  al.}}{{Galametz} et~al.}{2011}]{Galametz11}
{Galametz} M.,  {Madden} S.~C.,  {Galliano} F.,   {et al.} 2011, \mn@doi [\aap]
  {10.1051/0004-6361/201014904}, \href
  {http://adsabs.harvard.edu/abs/2011A%26A...532A..56G} {532, A56}

\bibitem[\protect\citeauthoryear{{Galliano}, {Dwek}  \& {Chanial}}{{Galliano}
  et~al.}{2008}]{Galliano08}
{Galliano} F.,  {Dwek} E.,   {Chanial} P.,  2008, \mn@doi [\apj]
  {10.1086/523621}, \href {http://adsabs.harvard.edu/abs/2008ApJ...672..214G}
  {672, 214}

\bibitem[\protect\citeauthoryear{{Galliano}, {Hony}, {Bernard}  \& {et
  al.}}{{Galliano} et~al.}{2011}]{Galliano11}
{Galliano} F.,  {Hony} S.,  {Bernard} J.-P.,   {et al.} 2011, \mn@doi [\aap]
  {10.1051/0004-6361/201117952}, \href
  {http://adsabs.harvard.edu/abs/2011A%26A...536A..88G} {536, A88}

\bibitem[\protect\citeauthoryear{{Gil de Paz}, {Boissier}, {Madore}  \& {et
  al.}}{{Gil de Paz} et~al.}{2007}]{GilDePaz07}
{Gil de Paz} A.,  {Boissier} S.,  {Madore} B.~F.,   {et al.} 2007, \mn@doi
  [\apjs] {10.1086/516636}, \href
  {http://adsabs.harvard.edu/abs/2007ApJS..173..185G} {173, 185}

\bibitem[\protect\citeauthoryear{{Gomez}, {Baes}, {Cortese}  \& {et
  al.}}{{Gomez} et~al.}{2010}]{Gomez10}
{Gomez} H.~L.,  {Baes} M.,  {Cortese} L.,   {et al.} 2010, \mn@doi [\aap]
  {10.1051/0004-6361/201014530}, \href
  {http://adsabs.harvard.edu/abs/2010A%26A...518L..45G} {518, L45}

\bibitem[\protect\citeauthoryear{{Griffin}, {Abergel}, {Abreu}  \& {et
  al.}}{{Griffin} et~al.}{2010}]{Griffin10}
{Griffin} M.~J.,  {Abergel} A.,  {Abreu} A.,   {et al.} 2010, \mn@doi [\aap]
  {10.1051/0004-6361/201014519}, \href
  {http://adsabs.harvard.edu/abs/2010A%26A...518L...3G} {518, L3}

\bibitem[\protect\citeauthoryear{{Hirashita}, {Nozawa}, {Villaume}  \&
  {Srinivasan}}{{Hirashita} et~al.}{2015}]{Hirashita15}
{Hirashita} H.,  {Nozawa} T.,  {Villaume} A.,   {Srinivasan} S.,  2015, \mn@doi
  [\mnras] {10.1093/mnras/stv2095}, \href
  {http://cdsads.u-strasbg.fr/abs/2015MNRAS.454.1620H} {454, 1620}

\bibitem[\protect\citeauthoryear{{Houck}, {Roellig}, {van Cleve}  \& {et
  al.}}{{Houck} et~al.}{2004}]{Houck04}
{Houck} J.~R.,  {Roellig} T.~L.,  {van Cleve} J.,   {et al.} 2004, \mn@doi
  [\apjs] {10.1086/423134}, \href
  {http://cdsads.u-strasbg.fr/abs/2004ApJS..154...18H} {154, 18}

\bibitem[\protect\citeauthoryear{{Indebetouw}, {Mathis}, {Babler}  \& {et
  al.}}{{Indebetouw} et~al.}{2005}]{Indebetouw05}
{Indebetouw} R.,  {Mathis} J.~S.,  {Babler} B.~L.,   {et al.} 2005, \mn@doi
  [\apj] {10.1086/426679}, \href
  {http://adsabs.harvard.edu/abs/2005ApJ...619..931I} {619, 931}

\bibitem[\protect\citeauthoryear{{Jarrett}, {Cohen}, {Masci}  \& {et
  al.}}{{Jarrett} et~al.}{2011}]{Jarrett11}
{Jarrett} T.~H.,  {Cohen} M.,  {Masci} F.,   {et al.} 2011, \mn@doi [\apj]
  {10.1088/0004-637X/735/2/112}, \href
  {http://adsabs.harvard.edu/abs/2011ApJ...735..112J} {735, 112}

\bibitem[\protect\citeauthoryear{{Johansson}, {Naab}  \&
  {Ostriker}}{{Johansson} et~al.}{2012}]{Johansson12}
{Johansson} P.~H.,  {Naab} T.,   {Ostriker} J.~P.,  2012, \mn@doi [\apj]
  {10.1088/0004-637X/754/2/115}, \href
  {http://adsabs.harvard.edu/abs/2012ApJ...754..115J} {754, 115}

\bibitem[\protect\citeauthoryear{{Kaneda}, {Onaka}, {Kitayama}, {Okada}  \&
  {Sakon}}{{Kaneda} et~al.}{2007}]{Kaneda07}
{Kaneda} H.,  {Onaka} T.,  {Kitayama} T.,  {Okada} Y.,   {Sakon} I.,  2007,
  \mn@doi [\pasj] {10.1093/pasj/59.1.107}, \href
  {http://cdsads.u-strasbg.fr/abs/2007PASJ...59..107K} {59, 107}

\bibitem[\protect\citeauthoryear{{Kaneda}, {Onaka}, {Sakon}, {Kitayama},
  {Okada}, {Suzuki}, {Ishihara}  \& {Yamagishi}}{{Kaneda}
  et~al.}{2010}]{Kaneda10}
{Kaneda} H.,  {Onaka} T.,  {Sakon} I.,  {Kitayama} T.,  {Okada} Y.,  {Suzuki}
  T.,  {Ishihara} D.,   {Yamagishi} M.,  2010, \mn@doi [\apjl]
  {10.1088/2041-8205/716/2/L161}, \href
  {http://cdsads.u-strasbg.fr/abs/2010ApJ...716L.161K} {716, L161}

\bibitem[\protect\citeauthoryear{{Kaviraj}}{{Kaviraj}}{2014a}]{Kaviraj14a}
{Kaviraj} S.,  2014a, \mn@doi [\mnras] {10.1093/mnrasl/slt136}, \href
  {http://adsabs.harvard.edu/abs/2014MNRAS.437L..41K} {437, L41}

\bibitem[\protect\citeauthoryear{{Kaviraj}}{{Kaviraj}}{2014b}]{Kaviraj14b}
{Kaviraj} S.,  2014b, \mn@doi [\mnras] {10.1093/mnras/stu338}, \href
  {http://adsabs.harvard.edu/abs/2014MNRAS.440.2944K} {440, 2944}

\bibitem[\protect\citeauthoryear{{Kaviraj}, {Tan}, {Ellis}  \&
  {Silk}}{{Kaviraj} et~al.}{2011}]{Kaviraj11}
{Kaviraj} S.,  {Tan} K.-M.,  {Ellis} R.~S.,   {Silk} J.,  2011, \mn@doi
  [\mnras] {10.1111/j.1365-2966.2010.17754.x}, \href
  {http://adsabs.harvard.edu/abs/2011MNRAS.411.2148K} {411, 2148}

\bibitem[\protect\citeauthoryear{{Kaviraj}, {Crockett}, {Whitmore}  \& {et
  al.}}{{Kaviraj} et~al.}{2012}]{Kaviraj12}
{Kaviraj} S.,  {Crockett} R.~M.,  {Whitmore} B.~C.,   {et al.} 2012, \mn@doi
  [\mnras] {10.1111/j.1745-3933.2012.01246.x}, \href
  {http://adsabs.harvard.edu/abs/2012MNRAS.422L..96K} {422, L96}

\bibitem[\protect\citeauthoryear{{Kaviraj}, {Rowlands}, {Alpaslan}  \& {et
  al.}}{{Kaviraj} et~al.}{2013}]{Kaviraj13}
{Kaviraj} S.,  {Rowlands} K.,  {Alpaslan} M.,   {et al.} 2013, \mn@doi [\mnras]
  {10.1093/mnras/stt1629}, \href
  {http://adsabs.harvard.edu/abs/2013MNRAS.435.1463K} {435, 1463}

\bibitem[\protect\citeauthoryear{{Kennicutt}}{{Kennicutt}}{1998}]{Kennicutt98a}
{Kennicutt} Jr. R.~C.,  1998, \mn@doi [\apj] {10.1086/305588}, \href
  {http://adsabs.harvard.edu/abs/1998ApJ...498..541K} {498, 541}

\bibitem[\protect\citeauthoryear{{Kennicutt}, {Calzetti}, {Aniano}  \& {et
  al.}}{{Kennicutt} et~al.}{2011}]{Kennicutt11}
{Kennicutt} R.~C.,  {Calzetti} D.,  {Aniano} G.,   {et al.} 2011, \mn@doi
  [\pasp] {10.1086/663818}, \href
  {http://cdsads.u-strasbg.fr/abs/2011PASP..123.1347K} {123, 1347}

\bibitem[\protect\citeauthoryear{{Kim}, {Sheth}, {Hinz}  \& {et al.}}{{Kim}
  et~al.}{2012}]{Kim12}
{Kim} T.,  {Sheth} K.,  {Hinz} J.~L.,   {et al.} 2012, \mn@doi [\apj]
  {10.1088/0004-637X/753/1/43}, \href
  {http://adsabs.harvard.edu/abs/2012ApJ...753...43K} {753, 43}

\bibitem[\protect\citeauthoryear{{Kulkarni}, {Sahu}, {Chaware}, {Chakradhari}
  \& {Pandey}}{{Kulkarni} et~al.}{2014}]{Kulkarni14}
{Kulkarni} S.,  {Sahu} D.~K.,  {Chaware} L.,  {Chakradhari} N.~K.,   {Pandey}
  S.~K.,  2014, \mn@doi [\na] {10.1016/j.newast.2014.01.003}, \href
  {http://adsabs.harvard.edu/abs/2014NewA...30...51K} {30, 51}

\bibitem[\protect\citeauthoryear{{Lackner}, {Cen}, {Ostriker}  \&
  {Joung}}{{Lackner} et~al.}{2012}]{Lackner12}
{Lackner} C.~N.,  {Cen} R.,  {Ostriker} J.~P.,   {Joung} M.~R.,  2012, \mn@doi
  [\mnras] {10.1111/j.1365-2966.2012.21525.x}, \href
  {http://cdsads.u-strasbg.fr/abs/2012MNRAS.425..641L} {425, 641}

\bibitem[\protect\citeauthoryear{{Lianou}, {Barmby}  \& {Taylor}}{{Lianou}
  et~al.}{2013}]{Lianou13}
{Lianou} S.,  {Barmby} P.,   {Taylor} J.,  2013, in 11th Hellenic Astronomical
  Conference. pp 31--31

\bibitem[\protect\citeauthoryear{{Lianou}, {Barmby}, {R{\'e}my-Ruyer},
  {Madden}, {Galliano}  \& {Lebouteiller}}{{Lianou} et~al.}{2014}]{Lianou14}
{Lianou} S.,  {Barmby} P.,  {R{\'e}my-Ruyer} A.,  {Madden} S.~C.,  {Galliano}
  F.,   {Lebouteiller} V.,  2014, \mn@doi [\mnras] {10.1093/mnras/stu1770},
  \href {http://adsabs.harvard.edu/abs/2014MNRAS.445.1003L} {445, 1003}

\bibitem[\protect\citeauthoryear{{Madden}, {R{\'e}my-Ruyer}, {Galametz}  \& {et
  al.}}{{Madden} et~al.}{2013}]{Madden13}
{Madden} S.~C.,  {R{\'e}my-Ruyer} A.,  {Galametz} M.,   {et al.} 2013, \mn@doi
  [\pasp] {10.1086/671138}, \href
  {http://cdsads.u-strasbg.fr/abs/2013PASP..125..600M} {125, 600}

\bibitem[\protect\citeauthoryear{{Magdis} et~al.,}{{Magdis}
  et~al.}{2012}]{Magdis12}
{Magdis} G.~E.,  et~al., 2012, \mn@doi [\apj] {10.1088/0004-637X/760/1/6},
  \href {http://cdsads.u-strasbg.fr/abs/2012ApJ...760....6M} {760, 6}

\bibitem[\protect\citeauthoryear{{Martig}, {Bournaud}, {Teyssier}  \&
  {Dekel}}{{Martig} et~al.}{2009}]{Martig09}
{Martig} M.,  {Bournaud} F.,  {Teyssier} R.,   {Dekel} A.,  2009, \mn@doi
  [\apj] {10.1088/0004-637X/707/1/250}, \href
  {http://adsabs.harvard.edu/abs/2009ApJ...707..250M} {707, 250}

\bibitem[\protect\citeauthoryear{{Martig}, {Crocker}, {Bournaud}  \& {et
  al.}}{{Martig} et~al.}{2013}]{Martig13}
{Martig} M.,  {Crocker} A.~F.,  {Bournaud} F.,   {et al.} 2013, \mn@doi
  [\mnras] {10.1093/mnras/sts594}, \href
  {http://adsabs.harvard.edu/abs/2013MNRAS.432.1914M} {432, 1914}

\bibitem[\protect\citeauthoryear{{Misiriotis}, {Xilouris}, {Papamastorakis},
  {Boumis}  \& {Goudis}}{{Misiriotis} et~al.}{2006}]{Misiriotis06}
{Misiriotis} A.,  {Xilouris} E.~M.,  {Papamastorakis} J.,  {Boumis} P.,
  {Goudis} C.~D.,  2006, \mn@doi [\aap] {10.1051/0004-6361:20054618}, \href
  {http://cdsads.u-strasbg.fr/abs/2006A%26A...459..113M} {459, 113}

\bibitem[\protect\citeauthoryear{{Moellenhoff}}{{Moellenhoff}}{1981}]{Moellenhoff81}
{Moellenhoff} C.,  1981, \aap, \href
  {http://adsabs.harvard.edu/abs/1981A%26A....99..341M} {99, 341}

\bibitem[\protect\citeauthoryear{{Morrissey}, {Conrow}, {Barlow}  \& {et
  al.}}{{Morrissey} et~al.}{2007}]{Morrissey07}
{Morrissey} P.,  {Conrow} T.,  {Barlow} T.~A.,   {et al.} 2007, \mn@doi [\apjs]
  {10.1086/520512}, \href {http://adsabs.harvard.edu/abs/2007ApJS..173..682M}
  {173, 682}

\bibitem[\protect\citeauthoryear{{Nenkova}, {Sirocky}, {Ivezi{\'c}}  \&
  {Elitzur}}{{Nenkova} et~al.}{2008}]{Nenkova08}
{Nenkova} M.,  {Sirocky} M.~M.,  {Ivezi{\'c}} {\v Z}.,   {Elitzur} M.,  2008,
  \mn@doi [\apj] {10.1086/590482}, \href
  {http://cdsads.u-strasbg.fr/abs/2008ApJ...685..147N} {685, 147}

\bibitem[\protect\citeauthoryear{{Noeske}, {Weiner}, {Faber}  \& {et
  al.}}{{Noeske} et~al.}{2007}]{Noeske07}
{Noeske} K.~G.,  {Weiner} B.~J.,  {Faber} S.~M.,   {et al.} 2007, \mn@doi
  [\apjl] {10.1086/517926}, \href
  {http://cdsads.u-strasbg.fr/abs/2007ApJ...660L..43N} {660, L43}

\bibitem[\protect\citeauthoryear{{Noll}, {Burgarella}, {Giovannoli}  \& {et
  al.}}{{Noll} et~al.}{2009}]{Noll09}
{Noll} S.,  {Burgarella} D.,  {Giovannoli} E.,   {et al.} 2009, \mn@doi [\aap]
  {10.1051/0004-6361/200912497}, \href
  {http://adsabs.harvard.edu/abs/2009A%26A...507.1793N} {507, 1793}

\bibitem[\protect\citeauthoryear{{Oser}, {Ostriker}, {Naab}, {Johansson}  \&
  {Burkert}}{{Oser} et~al.}{2010}]{Oser10}
{Oser} L.,  {Ostriker} J.~P.,  {Naab} T.,  {Johansson} P.~H.,   {Burkert} A.,
  2010, \mn@doi [\apj] {10.1088/0004-637X/725/2/2312}, \href
  {http://cdsads.u-strasbg.fr/abs/2010ApJ...725.2312O} {725, 2312}

\bibitem[\protect\citeauthoryear{{Padmanabhan}, {Schlegel}, {Finkbeiner}  \&
  {et al.}}{{Padmanabhan} et~al.}{2008}]{Padmanabhan08}
{Padmanabhan} N.,  {Schlegel} D.~J.,  {Finkbeiner} D.~P.,   {et al.} 2008,
  \mn@doi [\apj] {10.1086/524677}, \href
  {http://adsabs.harvard.edu/abs/2008ApJ...674.1217P} {674, 1217}

\bibitem[\protect\citeauthoryear{{Panuzzo}, {Rampazzo}, {Bressan}, {Vega},
  {Annibali}, {Buson}, {Clemens}  \& {Zeilinger}}{{Panuzzo}
  et~al.}{2011}]{Panuzzo11}
{Panuzzo} P.,  {Rampazzo} R.,  {Bressan} A.,  {Vega} O.,  {Annibali} F.,
  {Buson} L.~M.,  {Clemens} M.~S.,   {Zeilinger} W.~W.,  2011, \mn@doi [\aap]
  {10.1051/0004-6361/201015908}, \href
  {http://cdsads.u-strasbg.fr/abs/2011A%26A...528A..10P} {528, A10}

\bibitem[\protect\citeauthoryear{{Parkin}, {Wilson}, {Foyle}  \& {et
  al.}}{{Parkin} et~al.}{2012}]{Parkin12}
{Parkin} T.~J.,  {Wilson} C.~D.,  {Foyle} K.,   {et al.} 2012, \mn@doi [\mnras]
  {10.1111/j.1365-2966.2012.20778.x}, \href
  {http://adsabs.harvard.edu/abs/2012MNRAS.422.2291P} {422, 2291}

\bibitem[\protect\citeauthoryear{{Pastorello}, {Forbes}, {Foster}, {Brodie},
  {Usher}, {Romanowsky}, {Strader}  \& {Arnold}}{{Pastorello}
  et~al.}{2014}]{Pastorello14}
{Pastorello} N.,  {Forbes} D.~A.,  {Foster} C.,  {Brodie} J.~P.,  {Usher} C.,
  {Romanowsky} A.~J.,  {Strader} J.,   {Arnold} J.~A.,  2014, \mn@doi [\mnras]
  {10.1093/mnras/stu937}, \href
  {http://cdsads.u-strasbg.fr/abs/2014MNRAS.442.1003P} {442, 1003}

\bibitem[\protect\citeauthoryear{{Pilyugin} \& {Thuan}}{{Pilyugin} \&
  {Thuan}}{2005}]{Pilyugin05}
{Pilyugin} L.~S.,  {Thuan} T.~X.,  2005, \mn@doi [\apj] {10.1086/432408}, \href
  {http://adsabs.harvard.edu/abs/2005ApJ...631..231P} {631, 231}

\bibitem[\protect\citeauthoryear{{Poglitsch}, {Waelkens}, {Geis}  \& {et
  al.}}{{Poglitsch} et~al.}{2010}]{Poglitsch10}
{Poglitsch} A.,  {Waelkens} C.,  {Geis} N.,   {et al.} 2010, \mn@doi [\aap]
  {10.1051/0004-6361/201014535}, \href
  {http://adsabs.harvard.edu/abs/2010A%26A...518L...2P} {518, L2}

\bibitem[\protect\citeauthoryear{{R{\'e}my-Ruyer}, {Madden}, {Galliano}  \& {et
  al.}}{{R{\'e}my-Ruyer} et~al.}{2013}]{Remy13}
{R{\'e}my-Ruyer} A.,  {Madden} S.~C.,  {Galliano} F.,   {et al.} 2013, \mn@doi
  [\aap] {10.1051/0004-6361/201321602}, \href
  {http://adsabs.harvard.edu/abs/2013A%26A...557A..95R} {557, A95}

\bibitem[\protect\citeauthoryear{{R{\'e}my-Ruyer}, {Madden}, {Galliano}  \& {et
  al.}}{{R{\'e}my-Ruyer} et~al.}{2014}]{Remy14}
{R{\'e}my-Ruyer} A.,  {Madden} S.~C.,  {Galliano} F.,   {et al.} 2014, \mn@doi
  [\aap] {10.1051/0004-6361/201322803}, \href
  {http://adsabs.harvard.edu/abs/2014A%26A...563A..31R} {563, A31}

\bibitem[\protect\citeauthoryear{{Renzini}}{{Renzini}}{2006}]{Renzini06}
{Renzini} A.,  2006, \mn@doi [\araa] {10.1146/annurev.astro.44.051905.092450},
  \href {http://cdsads.u-strasbg.fr/abs/2006ARA%26A..44..141R} {44, 141}

\bibitem[\protect\citeauthoryear{{Rieke}, {Young}, {Engelbracht}  \& {et
  al.}}{{Rieke} et~al.}{2004}]{Rieke04}
{Rieke} G.~H.,  {Young} E.~T.,  {Engelbracht} C.~W.,   {et al.} 2004, \mn@doi
  [\apjs] {10.1086/422717}, \href
  {http://adsabs.harvard.edu/abs/2004ApJS..154...25R} {154, 25}

\bibitem[\protect\citeauthoryear{{Roehlly}, {Burgarella}, {Buat}, {Boquien},
  {Ciesla}  \& {Heinis}}{{Roehlly} et~al.}{2014}]{Roehlly14}
{Roehlly} Y.,  {Burgarella} D.,  {Buat} V.,  {Boquien} M.,  {Ciesla} L.,
  {Heinis} S.,  2014, in {Manset} N.,  {Forshay} P.,  eds,  Astronomical
  Society of the Pacific Conference Series Vol. 485, Astronomical Data Analysis
  Software and Systems XXIII. p.~347

\bibitem[\protect\citeauthoryear{{Rowlands} et~al.,}{{Rowlands}
  et~al.}{2012}]{Rowlands12}
{Rowlands} K.,  et~al., 2012, \mn@doi [\mnras]
  {10.1111/j.1365-2966.2011.19905.x}, \href
  {http://adsabs.harvard.edu/abs/2012MNRAS.419.2545R} {419, 2545}

\bibitem[\protect\citeauthoryear{{Salpeter}}{{Salpeter}}{1955}]{Salpeter55}
{Salpeter} E.~E.,  1955, \mn@doi [\apj] {10.1086/145971}, \href
  {http://cdsads.u-strasbg.fr/abs/1955ApJ...121..161S} {121, 161}

\bibitem[\protect\citeauthoryear{{Schreiber}, {Pannella}, {Elbaz}  \& {et
  al.}}{{Schreiber} et~al.}{2015}]{Schreiber15}
{Schreiber} C.,  {Pannella} M.,  {Elbaz} D.,   {et al.} 2015, \mn@doi [\aap]
  {10.1051/0004-6361/201425017}, \href
  {http://cdsads.u-strasbg.fr/abs/2015A%26A...575A..74S} {575, A74}

\bibitem[\protect\citeauthoryear{{Seibert}, {Martin}, {Heckman}  \& {et
  al.}}{{Seibert} et~al.}{2005}]{Seibert05}
{Seibert} M.,  {Martin} D.~C.,  {Heckman} T.~M.,   {et al.} 2005, \mn@doi
  [\apjl] {10.1086/427843}, \href
  {http://adsabs.harvard.edu/abs/2005ApJ...619L..55S} {619, L55}

\bibitem[\protect\citeauthoryear{{Serra}, {Oosterloo}, {Morganti}  \& {et
  al.}}{{Serra} et~al.}{2012}]{Serra12}
{Serra} P.,  {Oosterloo} T.,  {Morganti} R.,   {et al.} 2012, \mn@doi [\mnras]
  {10.1111/j.1365-2966.2012.20219.x}, \href
  {http://adsabs.harvard.edu/abs/2012MNRAS.422.1835S} {422, 1835}

\bibitem[\protect\citeauthoryear{{Skrutskie}, {Cutri}, {Stiening}  \& {et
  al.}}{{Skrutskie} et~al.}{2006}]{Skrutskie06}
{Skrutskie} M.~F.,  {Cutri} R.~M.,  {Stiening} R.,   {et al.} 2006, \mn@doi
  [\aj] {10.1086/498708}, \href
  {http://adsabs.harvard.edu/abs/2006AJ....131.1163S} {131, 1163}

\bibitem[\protect\citeauthoryear{{Smith}, {Gomez}, {Eales}  \& {et
  al.}}{{Smith} et~al.}{2012}]{Smith12}
{Smith} M.~W.~L.,  {Gomez} H.~L.,  {Eales} S.~A.,   {et al.} 2012, \mn@doi
  [\apj] {10.1088/0004-637X/748/2/123}, \href
  {http://adsabs.harvard.edu/abs/2012ApJ...748..123S} {748, 123}

\bibitem[\protect\citeauthoryear{{Temi}, {Brighenti}, {Mathews}  \&
  {Bregman}}{{Temi} et~al.}{2004}]{Temi04}
{Temi} P.,  {Brighenti} F.,  {Mathews} W.~G.,   {Bregman} J.~D.,  2004, \mn@doi
  [\apjs] {10.1086/381963}, \href
  {http://cdsads.u-strasbg.fr/abs/2004ApJS..151..237T} {151, 237}

\bibitem[\protect\citeauthoryear{{Thomas}, {Maraston}, {Schawinski}, {Sarzi}
  \& {Silk}}{{Thomas} et~al.}{2010}]{Thomas10}
{Thomas} D.,  {Maraston} C.,  {Schawinski} K.,  {Sarzi} M.,   {Silk} J.,  2010,
  \mn@doi [\mnras] {10.1111/j.1365-2966.2010.16427.x}, \href
  {http://adsabs.harvard.edu/abs/2010MNRAS.404.1775T} {404, 1775}

\bibitem[\protect\citeauthoryear{{Tremonti} et~al.,}{{Tremonti}
  et~al.}{2004}]{Tremonti04}
{Tremonti} C.~A.,  et~al., 2004, \mn@doi [\apj] {10.1086/423264}, \href
  {http://cdsads.u-strasbg.fr/abs/2004ApJ...613..898T} {613, 898}

\bibitem[\protect\citeauthoryear{{Walcher}, {Groves}, {Budav{\'a}ri}  \&
  {Dale}}{{Walcher} et~al.}{2011}]{Walcher11}
{Walcher} J.,  {Groves} B.,  {Budav{\'a}ri} T.,   {Dale} D.,  2011, \mn@doi
  [\apss] {10.1007/s10509-010-0458-z}, \href
  {http://cdsads.u-strasbg.fr/abs/2011Ap%26SS.331....1W} {331, 1}

\bibitem[\protect\citeauthoryear{{Werner} et~al.,}{{Werner}
  et~al.}{2014}]{Werner14}
{Werner} N.,  et~al., 2014, \mn@doi [\mnras] {10.1093/mnras/stu006}, \href
  {http://cdsads.u-strasbg.fr/abs/2014MNRAS.439.2291W} {439, 2291}

\bibitem[\protect\citeauthoryear{{Wilson}}{{Wilson}}{1995}]{Wilson95}
{Wilson} C.~D.,  1995, \mn@doi [\apjl] {10.1086/309615}, \href
  {http://cdsads.u-strasbg.fr/abs/1995ApJ...448L..97W} {448, L97}

\bibitem[\protect\citeauthoryear{{Wilson}, {Cridland}, {Foyle}  \& {et
  al.}}{{Wilson} et~al.}{2013}]{Wilson13}
{Wilson} C.~D.,  {Cridland} A.,  {Foyle} K.,   {et al.} 2013, \mn@doi [\apjl]
  {10.1088/2041-8205/776/2/L30}, \href
  {http://adsabs.harvard.edu/abs/2013ApJ...776L..30W} {776, L30}

\bibitem[\protect\citeauthoryear{{Wright}, {Eisenhardt}, {Mainzer}  \& {et
  al.}}{{Wright} et~al.}{2010}]{Wright10}
{Wright} E.~L.,  {Eisenhardt} P.~R.~M.,  {Mainzer} A.~K.,   {et al.} 2010,
  \mn@doi [\aj] {10.1088/0004-6256/140/6/1868}, \href
  {http://adsabs.harvard.edu/abs/2010AJ....140.1868W} {140, 1868}

\bibitem[\protect\citeauthoryear{{Xilouris}, {Byun}, {Kylafis}, {Paleologou}
  \& {Papamastorakis}}{{Xilouris} et~al.}{1999}]{Xilouris99}
{Xilouris} E.~M.,  {Byun} Y.~I.,  {Kylafis} N.~D.,  {Paleologou} E.~V.,
  {Papamastorakis} J.,  1999, \aap, \href
  {http://adsabs.harvard.edu/abs/1999A%26A...344..868X} {344, 868}

\bibitem[\protect\citeauthoryear{{Xilouris}, {Madden}, {Galliano}, {Vigroux}
  \& {Sauvage}}{{Xilouris} et~al.}{2004}]{Xilouris04}
{Xilouris} E.~M.,  {Madden} S.~C.,  {Galliano} F.,  {Vigroux} L.,   {Sauvage}
  M.,  2004, \mn@doi [\aap] {10.1051/0004-6361:20034020}, \href
  {http://adsabs.harvard.edu/abs/2004A%26A...416...41X} {416, 41}

\bibitem[\protect\citeauthoryear{{Yi} et~al.,}{{Yi} et~al.}{2005}]{Yi05}
{Yi} S.~K.,  et~al., 2005, \mn@doi [\apjl] {10.1086/422811}, \href
  {http://adsabs.harvard.edu/abs/2005ApJ...619L.111Y} {619, L111}

\bibitem[\protect\citeauthoryear{{York}, {Adelman}, {Anderson}  \& {et
  al.}}{{York} et~al.}{2000}]{York00}
{York} D.~G.,  {Adelman} J.,  {Anderson} Jr. J.~E.,   {et al.} 2000, \mn@doi
  [\aj] {10.1086/301513}, \href
  {http://adsabs.harvard.edu/abs/2000AJ....120.1579Y} {120, 1579}

\bibitem[\protect\citeauthoryear{{Young}, {Bureau}, {Davis}  \& {et
  al.}}{{Young} et~al.}{2011}]{Young11}
{Young} L.~M.,  {Bureau} M.,  {Davis} T.~A.,   {et al.} 2011, \mn@doi [\mnras]
  {10.1111/j.1365-2966.2011.18561.x}, \href
  {http://adsabs.harvard.edu/abs/2011MNRAS.414..940Y} {414, 940}

\bibitem[\protect\citeauthoryear{{Young}, {Scott}, {Serra}  \& {et
  al.}}{{Young} et~al.}{2014}]{Young14}
{Young} L.~M.,  {Scott} N.,  {Serra} P.,   {et al.} 2014, \mn@doi [\mnras]
  {10.1093/mnras/stt2474}, \href
  {http://cdsads.u-strasbg.fr/abs/2014MNRAS.444.3408Y} {444, 3408}

\bibitem[\protect\citeauthoryear{{Zinn} \& {West}}{{Zinn} \&
  {West}}{1984}]{Zinn84}
{Zinn} R.,  {West} M.~J.,  1984, \mn@doi [\apjs] {10.1086/190947}, \href
  {http://cdsads.u-strasbg.fr/abs/1984ApJS...55...45Z} {55, 45}

\bibitem[\protect\citeauthoryear{{da Cunha}, {Charlot}  \& {Elbaz}}{{da Cunha}
  et~al.}{2008}]{dacunha08}
{da Cunha} E.,  {Charlot} S.,   {Elbaz} D.,  2008, \mn@doi [\mnras]
  {10.1111/j.1365-2966.2008.13535.x}, \href
  {http://cdsads.u-strasbg.fr/abs/2008MNRAS.388.1595D} {388, 1595}

\bibitem[\protect\citeauthoryear{{da Cunha}, {Eminian}, {Charlot}  \&
  {Blaizot}}{{da Cunha} et~al.}{2010}]{dacunha10b}
{da Cunha} E.,  {Eminian} C.,  {Charlot} S.,   {Blaizot} J.,  2010, \mn@doi
  [\mnras] {10.1111/j.1365-2966.2010.16344.x}, \href
  {http://cdsads.u-strasbg.fr/abs/2010MNRAS.403.1894D} {403, 1894}

\bibitem[\protect\citeauthoryear{{de Looze}, {Baes}, {Bendo}  \& {et al.}}{{de
  Looze} et~al.}{2012}]{deLooze12}
{de Looze} I.,  {Baes} M.,  {Bendo} G.~J.,   {et al.} 2012, \mn@doi [\mnras]
  {10.1111/j.1365-2966.2012.22045.x}, \href
  {http://cdsads.u-strasbg.fr/abs/2012MNRAS.427.2797D} {427, 2797}

\makeatother
\end{thebibliography}

% Don't change these lines
\bsp	% typesetting comment
\label{lastpage}
\end{document}